%% file: main.tex
\documentclass[a4paper,conference]{IEEEtran}
\ifCLASSINFOpdf
\else
\fi
\ifCLASSOPTIONcompsoc
  \usepackage[caption=false,font=normalsize,labelfont=sf,textfont=sf]{subfig}
\else
  \usepackage[caption=false,font=footnotesize]{subfig}
\fi
\usepackage{hyperref}
\usepackage{doi}
\usepackage{url}
\usepackage{enumitem}

\usepackage[T1]{fontenc}
\usepackage{comment}
\usepackage{xcolor}
\usepackage{booktabs, tabularx}
\newcolumntype{Y}{>{\raggedright\arraybackslash}X}
\usepackage{colortbl}
\usepackage{here} 
\usepackage{multicol}
\usepackage{multirow}
\usepackage{todonotes}
\usepackage{float}

\usepackage{threeparttable}

\definecolor{myturquoise}{cmyk}{1.0, 0.167, 0.0, 0.6}
\definecolor{mycrimson}{cmyk}{0.0, 1.0, 0.67, 0.2}
\definecolor{myorange}{cmyk}{0.0, 0.4, 0.8 0.0}
\definecolor{myblue}{rgb}{0,.1,.4}
\definecolor{mygreen}{cmyk}{0.22, 0.0, 0.82, 0.31}
\definecolor{mycolor}{cmyk}{0.0, 0.4, 0.8 0.0}
\usepackage{amsmath,amssymb}

\usepackage{standalone}
\usepackage{tikz}
\usetikzlibrary{shapes,arrows, positioning, matrix, fit, decorations.markings}
\tikzstyle{vecArrow} = [thick, draw=gray, decoration={markings,mark=at position
   1 with {\arrow[semithick]{open triangle 60}}},
   double distance=1.4pt, shorten >= 5.5pt,
   preaction = {decorate},
   postaction = {draw,line width=1.4pt, white,shorten >= 4.5pt}]
\tikzstyle{innerWhite} = [semithick, draw=gray, white,line width=1.4pt, shorten >= 4.5pt]
\usepackage{pgf-umlsd}
\usepackage{ifthen}
\usepackage{tikz-3dplot}
\usetikzlibrary{decorations.pathreplacing}

\usetikzlibrary{calc}
\usepackage{relsize}

\tikzset{fontscale/.style = {font=\relsize{#1}}}

\usepackage{pgfplots}
\usepgfplotslibrary{fillbetween}

\begin{document}
%
\title{Improved anomaly detection by training an autoencoder with skip connections on images corrupted with Stain-shaped noise}

\author{\IEEEauthorblockN{Anne-Sophie Collin and Christophe De Vleeschouwer}
\IEEEauthorblockA{ICTEAM Institute, UCLouvain \\ 
Louvain-La-Neuve, Belgium \\
\{anne-sophie.collin, christophe.devleeschouwer\}@uclouvain.be}}


\maketitle

\begin{abstract}
In industrial vision, the anomaly detection problem can be addressed with an autoencoder trained to map an arbitrary image, i.e. with or without any defect, to a clean image, i.e. without any defect. In this approach, anomaly detection relies conventionally on the reconstruction residual or, alternatively, on the reconstruction uncertainty. To improve the sharpness of the reconstruction, we consider an autoencoder architecture with skip connections. In the common scenario where only clean images are available for training, we propose to corrupt them with a synthetic noise model to prevent the convergence of the network towards the identity mapping, and introduce an original Stain noise model for that purpose. We show that this model favors the reconstruction of clean images from arbitrary real-world images, regardless of the actual defects appearance. In addition to demonstrating the relevance of our approach, our validation provides the first consistent assessment of reconstruction-based methods, by comparing their performance over the MVTec AD dataset \cite{Bergmann2019}, both for pixel- and image-wise anomaly detection. Our implementation is available at \url{https://github.com/anncollin/AnomalyDetection-Keras}.


\end{abstract}


%
\IEEEpeerreviewmaketitle

\section{Introduction}
\label{sec:intro}
\input{Chapters/Introduction.tex}

\section{Related Work}
\label{sec:relatedWork}
\input{Chapters/RelatedWork.tex}

\section{Methods}
\label{sec:methods}
\input{Chapters/Methods.tex}

\section{Results}
\label{sec:results}
\input{Chapters/Results.tex}

\section{Conclusion}
\label{sec:conclusion}
\input{Chapters/Conclusion.tex}

\bibliographystyle{IEEEtran}
\bibliography{library}


\end{document}

%% file: Chapters/Introduction.tex
\begin{figure}[!t]
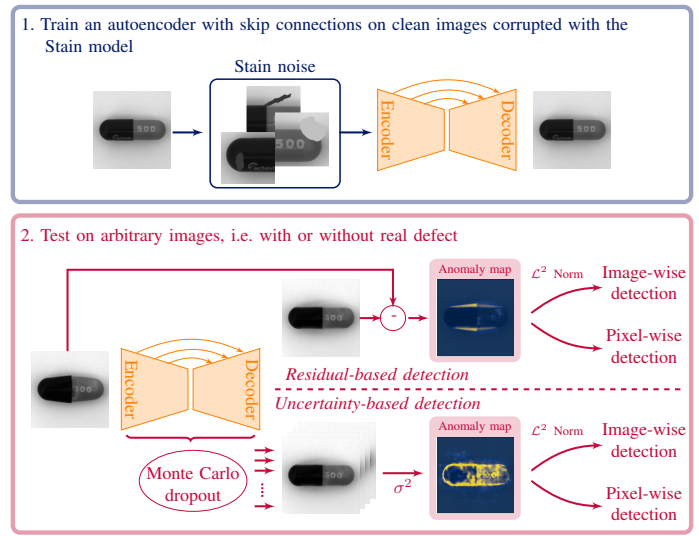

  \hspace*{-4.4cm}
  \includestandalone[width=0.75\textwidth]{imgs/introDiagram/tikz_stainOnly_vertical}
  \vspace*{-0.3cm}
  \caption{We improve the quality of the reconstructed images by training an autoencoder with skip connections over corrupted images. \textit{1. Blue block. } Corrupting the training images with our Stain noise model avoids the convergence of the network towards an unwanted identity operator. \textit{2. Red block.} The two anomaly detection strategies. In the upper part, the anomaly map is generated by subtracting the input image from its reconstruction. In the lower part, the anomaly map is estimated by the variance between 30 reconstructions inferred with Monte Carlo dropout (MCDropout) \cite{Kendall2017}. It relies on the hypothesis that structures that are not seen during training (defective areas) correlate with higher reconstruction uncertainty.}
  \label{fig:introDiagram}
\end{figure}

Anomaly detection can be defined as the task of identifying all diverging samples that does not belong to the distribution of regular, also named clean, data. When considering the specific application of the visual inspection of production lines, we are interested in detecting all defective samples occurring due to an unexpected behavior of the manufacturing process. This anomaly detection task could be formulated as a supervised learning problem. Such an approach uses both clean and defective examples to learn how to distinguish these two classes or even to refine the classification of defective samples into a variety of subclasses. However, the scarcity and variability of the defective samples make the data collection challenging and frequently produce unbalanced datasets \cite{Pimentel2014}. To circumvent the above-mentioned issues, anomaly detection is often formulated as an unsupervised learning task. This formulation makes it possible to either solve the detection problem itself or to ease the data collection process required by a supervised approach. \\
The unsupervised anomaly detection framework considered in this work is depicted in Figure \ref{fig:introDiagram}. It builds on the training of an autoencoder to project an arbitrary image onto the clean distribution of images (blue block). The training set is constituted exclusively of clean images. Then, defective structures can be inferred from the reconstruction (red block), following a traditional approach based on the residual \cite{Pimentel2014}, or even from an estimation of the prediction uncertainty \cite{Seebock2019}.\\

In conventional reconstruction-based approach, an autoencoder is trained on clean images only to perform an identity mapping. The bottleneck forces the network to learn a compressed representation of the training data that is expected to regularize the reconstruction towards the normal class. In the literature, the use of the Mean Squared Error (MSE) loss to train an hourglass CNN, without skip connections, has been criticized for its trend to produce blurry output images \cite{Zhao2015,Bergmann2019a}. Since the anomaly detection is based on the reconstruction residual, this behavior is detrimental because it alters the clean structures of an image as well as the defective ones.\\
A lot of effort has been made to improve the quality of the reconstructed images by the introduction of new loss functions. In this spirit, unsupervised methods based on Generative Adversarial Networks (GANs) have emerged \cite{Sabokrou2018,Schlegl2017,Schlegl2019,Akcay2019,Baur2019}. If GANs are known for their ability to produce realistic high-quality synthetic images \cite{Radford2015}, they have major drawbacks. Usually, GANs are difficult to train due to their trend to converge towards mode collapse \cite{Goodfellow1976}. In the context of anomaly detection, some GAN-based solutions fail to exclude defective samples from the generative distribution \cite{Akcay2019} and require an extra optimization step in the latent space during inference \cite{Schlegl2017}. Performances of AnoGAN \cite{Schlegl2017} over the MVTec AD dataset have been reported by Bergmann et al. \cite{Bergmann2019}. Those are significantly lower than the method proposed in this work. \\
Also, the use of a loss based on the Structural SIMilarity index has been considered for image generation, motivated by its ability to produce well looking images from a human perceptual perspective \cite{Zhao2015,Snell2017}. The SSIM have shown some improvement over the MSE loss for the training of an autoencoder in the context of anomaly detection. However, the SSIM loss formulation does not generalize to color images and is parametric. Traditionally, these hyper-parameters are tuned based on a validation set. However, in a real-life scenarios of anomaly detection, samples with real defects are usually not available. For this reason, our paper focus on the MSE rather than on the parametric SSIM.\\

With the objective of building our method on the MSE loss for its simplicity and widespread usage, we propose a new non-parametric approach that addresses the above-mentioned issues. To enhance the sharpness of the reconstruction, we consider an autoencoder equipped with skip connections, which allow the information to bypass the bottleneck. In order to prevent systematic transmission of the image structures through these links, the network is trained to reconstruct a clean image out of a corrupted version of it. As discussed later, the methodology used to corrupt the training images has a huge impact on the overall performances. We introduce a new synthetic model, named Stain, that adds an irregular elliptic structure of variable color and size to the input image.  Despite its simplicity, the Stain model is by far the best performing compared to the scene-specific corruption investigated in a previous study \cite{Huang2019}. Our Stain model has the double advantage of performing consistently better, while being independent of the image content.  \\ 


In Section \ref{sec:relatedWork} we provide an overview of previous reconstruction-based methods addressing the anomaly detection problem. Details of our method, including network architecture and the Stain noise model description, are provided in Section \ref{sec:methods}. In Section \ref{sec:results} we provide a comparative study of residual- and uncertainty-based anomaly detection strategies, both at the image and pixel level. To the best of our knowledge, our work is the first to provide a fair comparison (using the same dataset, comparable network architectures, covering a large variety of use cases) between the various detection strategies proposed in the recent literature. This extensive comparative study demonstrate the benefit of our proposed framework, combining skip connections and our novel corruption model.

%% file: imgs/introDiagram/tikz_stainOnly_vertical.tex
\tikzstyle{blueblock} = [draw=myblue, rectangle, rounded corners, minimum height=3em, minimum width=6em, line width=0.3mm]
\tikzstyle{boldblueblock} = [draw=myblue, rectangle, rounded corners, minimum height=3em, minimum width=6em, line width=0.85mm, opacity=0.4]
\tikzstyle{redblock} = [draw=mycrimson, rectangle, rounded corners, minimum height=3em, minimum width=6em]
\tikzstyle{boldredblock} = [draw=mycrimson, rectangle, rounded corners, minimum height=3em, minimum width=6em, line width=0.85mm, opacity=0.4]
\tikzstyle{anomalyblock} = [fill=mycrimson, rectangle, rounded corners, line width=0.7mm,opacity=0.2]
\tikzstyle{sum} = [draw=mycrimson, fill=white, circle]

\begin{tikzpicture}[node distance=2cm,>=latex', every text node part/.style={align=center}]

	\newcommand\mysize{0.4}
	\newcommand\tinysize{0.05}
	\newcommand\xx{-3.0}
	\newcommand\blocshift{3.0}   
	\newcommand\blocshifty{1.4} 


\begin{scope}[yshift= -2.3 cm]
\node at (2.5,2.15) [boldblueblock, text width=13cm, text height=3.6cm] (training) {};
\node at (-5.1,3.7) [fontscale=\mysize, text width=14cm, anchor=west] {\textcolor{myblue}{1. Train an autoencoder with skip connections on clean images corrupted with the }};
\node at (4.5,3.3) [fontscale=\mysize, text width=14cm, anchor=east] {\textcolor{myblue}{Stain model}};

\node at (-1.8,1.65) [fontscale=\mysize, text width=1.5cm] (input) {\includegraphics[width=1.5cm]{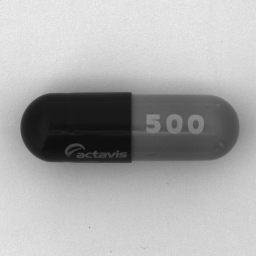}};

\node at (6.8,1.65) [text width=1.5cm] (out) {\includegraphics[width=1.5cm]{imgs/introDiagram/clean.png}};

	\draw[->, line width=0.4mm, color = myblue, yshift=2.1cm]
	(-1.0,-0.5) edge [solid] (-0.4,-0.5)
	(2.25,-0.5) to [solid] (2.9,-0.5);

	\begin{scope}[xshift= 1.0 cm, yshift= -0.1 cm]
	\node at (0,1.68) [blueblock,fontscale=\mysize, text width=2.3cm, text height=1.90cm] (corruption) {};
	\node at (0,1.65) {\includegraphics[width=2.1cm]{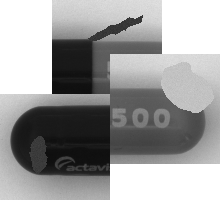}};
	\node at (0,3.0) [fontscale=\mysize,color=myblue] {Stain noise}; 
	\end{scope}
	
	\begin{scope}[xshift= 3.0 cm, yshift= -0.45 cm]
	
	\draw[fill=orange!60,fill opacity=0.4,draw=orange,yshift=2.1cm] (0,0.8) -- (0,-0.8) --(1.3,-0.4) -- (1.3,0.4) -- cycle; 
	\draw[fill=orange!60,fill opacity=0.4,draw=orange,yshift=2.1cm] (1.4,-0.4) -- (1.4,0.4) --(2.7,0.8) -- (2.7,-0.8) -- cycle;

	\draw[->,line width=0.2mm, color = orange,yshift=2.1cm]
	(0.3,0.72) edge [bend left=25,solid] (2.4,0.72)
	(0.65,0.6) edge [bend left=25,solid] (2.1, 0.6)
	(0.9,0.5) to [bend left=25,solid] (1.8, 0.5);

	\node at (0.2, 2.1) [fontscale=\mysize,rotate=270,color=orange] {Encoder}; 
	\node at (2.55, 2.1) [fontscale=\mysize,rotate=270,color=orange] {Decoder};
	\end{scope}
\end{scope}

	
\begin{scope}[yshift= -5.75cm]

\node at (2.5,0.3) [boldredblock, text width=13cm, text height=6.0cm] (training) {};
\node at (0.3,3.0) [fontscale=\mysize] {\textcolor{mycrimson}{2. Test on arbitrary images, i.e. with or without real defect}};

\node at (-3.0,0) [fontscale=\mysize, text width=1.5cm] (input) {\includegraphics[width=1.5cm]{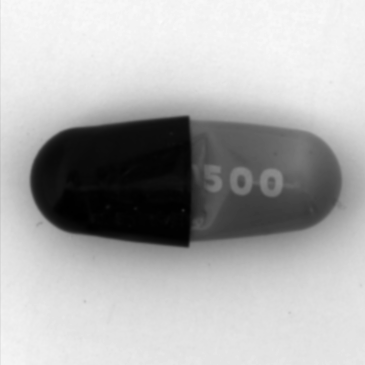}};

	\begin{scope}[xshift= -2.0 cm, yshift= -2.1cm]
	
	\draw[fill=orange!60,fill opacity=0.4,draw=orange,yshift=2.1cm] (0,0.8) -- (0,-0.8) --(1.3,-0.4) -- (1.3,0.4) -- cycle; 
	\draw[fill=orange!60,fill opacity=0.4,draw=orange,yshift=2.1cm] (1.4,-0.4) -- (1.4,0.4) --(2.7,0.8) -- (2.7,-0.8) -- cycle;
	
	\draw[->,line width=0.2mm, color = orange,yshift=2.1cm]
	(0.3,0.72) edge [bend left=25,solid] (2.4,0.72)
	(0.65,0.6) edge [bend left=25,solid] (2.1, 0.6)
	(0.9,0.5) to [bend left=25,solid] (1.8, 0.5);

	\node at (0.2, 2.1) [fontscale=\mysize,rotate=270,color=orange] {Encoder}; 
	\node at (2.55, 2.1) [fontscale=\mysize,rotate=270,color=orange] {Decoder};
	\end{scope}
	
\begin{scope}[xshift= 0.2cm,yshift= 1.4cm]

\node at (1.7,0) [text width=1.5cm] (out) {\includegraphics[width=1.5cm]{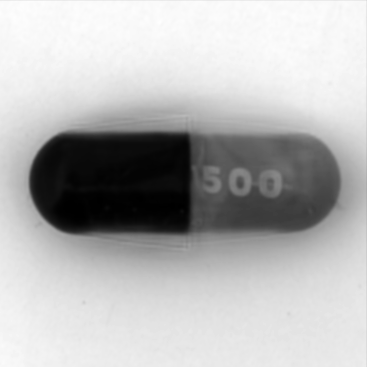}};

\node at (3.1,0) [sum,text width=0.2cm] (diff) {};
\node at (3.12,0) {\textcolor{mycrimson}{-}};

\node at (4.7,0.15) [anomalyblock,fontscale=\mysize, text width=1.55cm, text height=1.8cm] {};
\node at (4.7,0.15)[yshift=0.8cm]{\scriptsize \textcolor{mycrimson}{Anomaly map}};
\node at (4.7,0) [text width=1.5cm] (outvar) {\includegraphics[width=1.5cm]{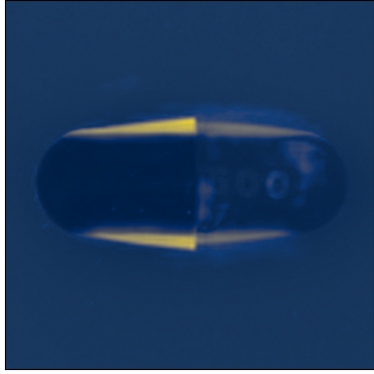}};
\draw[->, line width=0.4mm, color = mycrimson]
	(-3.25,-0.65) -- (-3.25,0.95) --  (3.1,0.95) -- (3.1,0.2);
\draw[->, line width=0.4mm, color = mycrimson]
	(2.45,0) edge (2.9,0)
	(3.4,0) to (3.8,0);

\draw[->, line width=0.4mm, color = mycrimson]
	(5.8,-0.1) edge [bend right=18,solid] (7.2,-0.6)
	(5.8,0.1) to [bend left=18,solid] (7.2,0.6);
\node at (8.0,-0.55) [text width=2.3cm,fontscale=\mysize] (det11) {\textcolor{mycrimson}{Pixel-wise detection}};
\node at (6.3,0.9) [text width=1.5cm] (l11) {\scriptsize\textcolor{mycrimson}{$\mathcal{L}^2$ Norm}};
\node at (8.0,0.7) [text width=2.5cm, fontscale=\mysize] (det11) {\textcolor{mycrimson}{Image-wise detection}};

\end{scope}

\draw[line width=0.4mm, color = mycrimson, dashed] (1,0) -- (9.0,0);
	\node at (3,0.3) [fontscale=\mysize] (app1) {\textcolor{mycrimson}{\textit{Residual-based detection}}};
	\node at (3,-0.3) [fontscale=\mysize] (app1) {\textcolor{mycrimson}{\textit{Uncertainty-based detection}}};

\begin{scope}[xshift= 0.2cm, yshift= -1.7cm]

\node at (1.7+0.3,0.3) [text width=1.5cm] (out) {\includegraphics[width=1.5cm]{imgs/introDiagram/Real_90_OUTPUT.png}};
\node at (1.7+0.2,0.2) [text width=1.5cm] (out) {\includegraphics[width=1.5cm]{imgs/introDiagram/Real_90_OUTPUT.png}};
\node at (1.7+0.1,0.1) [text width=1.5cm] (out) {\includegraphics[width=1.5cm]{imgs/introDiagram/Real_90_OUTPUT.png}};
\node at (1.7,0) [text width=1.5cm] (out) {\includegraphics[width=1.5cm]{imgs/introDiagram/Real_90_OUTPUT.png}};

\node at (4.7,0.15) [anomalyblock,fontscale=\mysize, text width=1.55cm, text height=1.8cm] {};
\node at (4.7,0.15)[yshift=0.8cm]{\scriptsize \textcolor{mycrimson}{Anomaly map}};
\node at (4.7,0) [text width=1.5cm] (outvar) {\includegraphics[width=1.5cm]{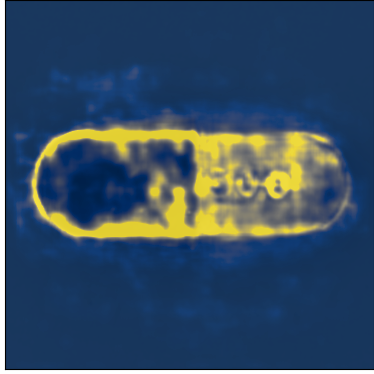}};

\draw[decorate,decoration={brace,amplitude=3pt,mirror}, line width=0.4mm, color = mycrimson] 
    (-2.1, 0.8) node(t_k_unten){} -- (0.4, 0.8) node(t_k_opt_unten){}; 
\node at (-0.8, -0.2) [fontscale=\mysize, text width=3.0cm] {\textcolor{mycrimson}{Monte Carlo dropout}};
\draw (-0.8, -0.15) [color= mycrimson,yshift=0.05cm] ellipse (1.05cm and 0.6cm);

\draw[->, line width=0.4mm, color = mycrimson]
	(0.4, 0.35+0.15) edge (0.8, 0.35+0.15)
	(0.4, 0.35-0.05) edge (0.8, 0.35-0.05)
	(0.4, 0.35-0.25) edge (0.8, 0.35-0.25)
	(0.4, 0.35-0.95) edge (0.8, 0.35-0.95)
	(2.9, 0.05) to (3.7, 0.05);
	\draw[line width=0.4mm, densely dotted,color = mycrimson]
	(0.55, -0.15) -- (0.55, -0.45);
	\node at (3.3, -0.2) [fontscale=\mysize, text width=3.0cm] {\textcolor{mycrimson}{$\sigma^2$}};

\draw[->, line width=0.4mm, color = mycrimson]
	(5.8,-0.1) edge [bend right=18,solid] (7.2,-0.6)
	(5.8,0.1) to [bend left=18,solid] (7.2,0.6);
\node at (8.0,-0.55) [text width=2.3cm,fontscale=\mysize] (det11) {\textcolor{mycrimson}{Pixel-wise detection}};
\node at (6.3,0.9) [text width=1.5cm] (l11) {\scriptsize\textcolor{mycrimson}{$\mathcal{L}^2$ Norm}};
\node at (8.0,0.7) [text width=2.5cm, fontscale=\mysize] (det11) {\textcolor{mycrimson}{Image-wise detection}};
\end{scope}
	
\end{scope}

\end{tikzpicture}

%% file: Chapters/RelatedWork.tex
Anomaly detection is a long-standing problem that has been considered in a variety of fields \cite{Pimentel2014,Chandola2009} and the reconstruction-based approach is one popular way to address the issue. In comparison to other methods for which the detection of abnormal samples is performed in another domain than the image \cite{Napoletano2018, Staar2019, Zhou2017}, reconstruction-based approaches offer the opportunity to identify the pixels that lead to the rejection of the image from the normal class. \\

Conventional reconstruction-based methods infer anomaly based on the reconstruction error between an arbitrary input and its reconstructed version. It assumes that clean structures are perfectly conserved while defective ones are replaced by clean content. However, when a defect contrasts poorly with its surroundings, replacing abnormal structures with clean content does not lead to a sufficiently high reconstruction error. In such cases, this methodology reaches the limit of its underlying assumptions. A previous study \cite{Seebock2019} detected anomalies by quantifying the prediction uncertainty with MCDropout \cite{Kendall2017} instead of the reconstruction residual. \\

To obtain a clean reconstruction out of an arbitrary image, an autoencoder, without skip connections, is generally trained to perform an image-to-image identity mapping with clean data only under the minimization of the MSE. This loss has the disadvantage of promoting blurry reconstructions, resulting in higher residuals for clean structures.\\
To improve the sharpness of the reconstruction, Bergmann et al. proposed a loss derived from the SSIM index \cite{Bergmann2019a}. However, the SSIM imposes to consider grayscale images and depends on multiple hyper-parameters, thereby hampering the reproducibility of the results.\\
Also, GANs have been considered to sample the clean distribution of images \cite{Sabokrou2018,Schlegl2017,Schlegl2019,Akcay2019,Baur2019}. Unfortunately, GANs are challenging to train due to their trend to converge towards mode collapse. Furthermore, the difficulty to exclude abnormal samples from the generative distribution penalizes performances \cite{Akcay2019}. Finally, some GAN-based methods require an optimization process during inference to find the latent space vector producing the most similar image between an arbitrary query and an output image belonging to the generative distribution \cite{Schlegl2017}. \\

Excluding defective structures from the distribution of generated images is a recurrent problem in anomaly detection. It is usually expected that the compression induced by the bottleneck is sufficient to regularize the reconstruction so that it lies on the clean images manifold. In practice, the autoencoder is not explicitly constrained to not reproduce abnormal content and often reconstructs defective structures. As an extension to traditional autoencoder-based methods studied in this work, a recent method proposed to mitigate this issue by iteratively projecting the arbitrary input towards the clean distribution of images. The projection is constrained to be similar, in the sense of the $\mathcal{L}^1$ norm, to the initial input \cite{Dehaene2020}. Instead of performing this optimization in the latent space as made with AnoGAN \cite{Schlegl2017}, they propose to find an optimal clean input image. If this practice enhances the sharpness of the reconstruction, the optimization step is resource consuming. \\
Also, the reconstruction task can be formulated as an image completion problem \cite{Haselmann2019,Munawar2015}. To make the inference and training phases consistent, it is assumed that the defects are entirely contained in the mask during inference, which limits the practical usage of the method. Mei et al. \cite{Mei2018} also proposed to use a denoising autoencoder to reconstruct training images corrupted with salt-and-pepper noise. However they did not discuss the gain brought by this modification, and only considered it for an hourglass CNN, without skip connections. \\

The methodology proposed in this work presents a simple approach to enhance the sharpness of the reconstruction. The skip connections allow the preservation of high frequency information by bypassing the bottleneck. However, we show that this practice penalizes anomaly detection when the model is trained to perform identity mapping on uncorrupted clean images. Nevertheless, the introduction of an original noise model allows to significantly improve the anomaly detection accuracy for the skipped architecture, which eventually outperforms the conventional one in many real-life cases.


%% file: Chapters/Methods.tex
Our method performs anomaly detection based on the regularized reconstruction predicted by an autoencoder. This section presents the different components of our approach, ranging from the training of the autoencoder to the strategies considered to detect defects based on the reconstruction residual or the reconstruction uncertainty.

\subsection{Model configuration} 
\begin{figure}[!t]
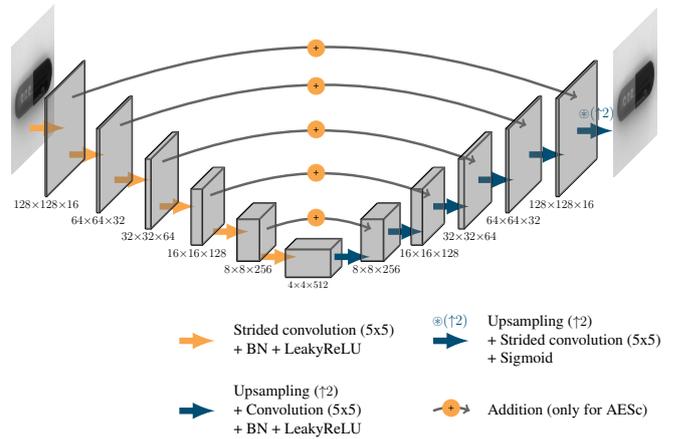

  \centering
  \includestandalone[width=0.48\textwidth]{imgs/Architecture/tikz_architecture_vertical}
  \caption{AESc architecture performing the projection of an arbitrary $256 \times 256$ image towards the distribution of clean images with the same dimension. Note that the AE architecture shares the same specifications with the exception of the skip connections that have been removed.}
  \label{fig:architecture}
\end{figure}

The reconstruction is based on a convolutional neural network. Our architecture, referred to as \textbf{Autoencoder with Skip-connections (AESc)} and shown in Figure \ref{fig:architecture}, is a variant of U-Net \cite{Ronneberger2015}. AESc takes input images of size $256 \times 256$ and projects them onto a latent space of $4 \times 4 \times 512$ dimension. The projection towards the lower dimensional space is performed by 6 consecutive convolutional layers strided by a factor 2. The back projection is performed by 6 layers of convolution followed by an upsampling operator of factor 2. All convolutions have a $5 \times 5$ kernel. Unlike the original U-Net version, our skip-connections perform an addition, not a concatenation, of feature maps of the encoder to the decoder. \\
For the sake of comparison, we also consider the \textbf{Autoencoder (AE)} network which follows exactly the same architecture but from which we removed the skip connections. 

\subsection{Corruption model} 

\begin{figure}[!t]
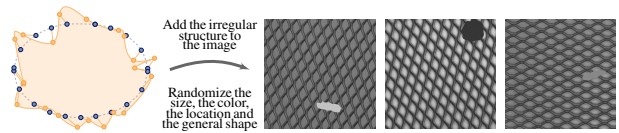

  \centering
  \includestandalone[width=0.45\textwidth]{imgs/StainModel/ellipse}
  \caption{The Stain noise model is a cubic interpolation between 20 points (orange dots), arranged in ascending order of polar coordinates, located around the border of an ellipse of variable size (blue line). The axes of the ellipse are comprised between 1 and 12\% of the smallest image dimension and its eccentricity is randomly initialized. }
  \label{fig:stainModel}
\end{figure}

Ideally, the autoencoder should preserve clean structures while modifying those that are not. Due to the impossibility of collecting pairs of clean and defective versions of the same sample, we propose to introduce synthetic corruption during training to explicitly constrain the autoencoder to remove this additive noise. Our \textbf{Stain} model, illustrated in Figure \ref{fig:introDiagram} and explained in Figure \ref{fig:stainModel}, corrupts images by adding a structure whose color is randomly selected in the grayscale range and whose shape is an ellipse with irregular edges. \\

The intuition behind the definition of this noise model is that occluding large area of the training images is a data augmentation procedure that helps to improve the network training \cite{Zhong2020, Fong2019}. Due to the skip-connections in our network architecture, this form of data augmentation is essential to avoid the convergence of the model towards the identity operator. However, we noticed that the use of regular shapes, like a conventional ellipse, leads to overfitting to this additive noise structure, as also pointed out in a context of inpainting with rectangular holes \cite{Liu2018}.

\subsection{Anomaly detection strategies} 
We compare two approaches to obtain the anomaly map representing the likelihood that a pixel is abnormal. On the one hand, the \textbf{residual-based} approach evaluates the abnormality by measuring the absolute difference between the input image $\mathbf{x}$ and its reconstruction $\mathbf{\hat{x}}$. On the other hand, the \textbf{uncertainty-based} approach relies on the intuition that structures that are not seen during training, i.e. the anomalies, will correlate with higher uncertainties, as estimated by the variance between 30 output images inferred with the  MCDropout technique. Our experiments revealed that more accurate detection is obtained by applying an increasing level of dropout for deepest layers. More specifically, the dropout levels are [0, 0, 10, 20, 30, 40] percent for layers ranging from the highest spatial resolution to the lowest. \\

Out of the anomaly map, it is either possible to classify the entire image as clean/defective or to classify each pixel as belonging to a clean/defective structure. In the first case, referred to as \textbf{image-wise detection}, it is common to compute the $\mathcal{L}^p$ norm of the anomaly map given by 
\begin{equation}
\mathcal{L}^p (\mathbf{x}, \mathbf{\hat{x}}) = \left( \sum_{i=0}^{m} \sum_{j=0}^{n} | \mathbf{x}_{i,j} - \mathbf{\hat{x}}_{i,j} |^p \right)^{1/p}
\end{equation}  
with $\mathbf{x}_{i,j}$ denoting the pixel belonging to the $i^{\text{th}}$ row and the $j^{\text{th}}$ column of the image $\mathbf{x}$ of size $m\times n$. Based on our experiments, we present results obtained for $p=2$ since they achieve the most stable accuracy values across the experiments. Hence, all images for which the $\mathcal{L}^2$ norm of the abnormality map exceeds a chosen threshold are considered as defective. In the second case, referred to as \textbf{pixel-wise detection}, the threshold is applied directly on each pixel value of the anomaly map. \\

To perform image-wise or pixel-wise anomaly detection, a threshold has to be determined. Since this threshold value is highly dependent on the application, we present the performances in terms of Area Under the receiver operating characteristic Curve (AUC), obtained by varying over the full range of threshold values.

%% file: imgs/Architecture/tikz_architecture_vertical.tex
\tikzstyle{graysum} = [draw=myorange, fill=myorange!100, circle, node distance=0.2cm]

\begin{tikzpicture}
\newcommand{\networkLayer}[7]{
\def\a{#1} 
\def\b{0.02}
\def\c{#2} 
\def\t{#3} 
\def\d{#4} 
\coordinate (#7) at (\t+\c/2,\a/2,\d+\a/2);
\draw[line width=0.5mm](\c+\t,0,\d) -- (\c+\t,\a,\d) -- (\t,\a,\d);                                                      
\draw[line width=0.5mm](\t,0,\a+\d) -- (\c+\t,0,\a+\d) node[midway,below] {#6} -- (\c+\t,\a,\a+\d) -- (\t,\a,\a+\d) -- (\t,0,\a+\d); 
\draw[line width=0.5mm](\c+\t,0,\d) -- (\c+\t,0,\a+\d);
\draw[line width=0.5mm](\c+\t,\a,\d) -- (\c+\t,\a,\a+\d);
\draw[line width=0.5mm](\t,\a,\d) -- (\t,\a,\a+\d);
\filldraw[#5] (\t+\b,\b,\a+\d) -- (\c+\t-\b,\b,\a+\d) -- (\c+\t-\b,\a-\b,\a+\d) -- (\t+\b,\a-\b,\a+\d) -- (\t+\b,\b,\a+\d); 
\filldraw[#5] (\t+\b,\a,\a-\b+\d) -- (\c+\t-\b,\a,\a-\b+\d) -- (\c+\t-\b,\a,\b+\d) -- (\t+\b,\a,\b+\d);
\ifthenelse {\equal{#5} {}}
{} 
{\filldraw[#5] (\c+\t,\b,\a-\b+\d) -- (\c+\t,\b,\b+\d) -- (\c+\t,\a-\b,\b+\d) -- (\c+\t,\a-\b,\a-\b+\d);} 
}

		\newcommand\x{1.3}
		\node[canvas is zy plane at x = 0.2, transform shape,anchor=south] at (-2.0,0.1){\includegraphics[width=4cm,height=4.5cm]{imgs/introDiagram/clean.png}};
		
		\newcommand\bending{20}
		\draw [->,color=black!60,line width=0.9mm] 
		(6.5,-2.5) edge [bend left=\bending] (10.2,-2.5)
		(4.5,-1.3) edge [bend left=\bending] (12.3,-1.3)
		(2.8,-0.1) edge [bend left=\bending] (13.9,-0.1)
		(1.3,1.2) edge [bend left=\bending] (15.7,1.2)
		(-0.4,2.2) to [bend left=\bending] (17.5,2.2);
		\node at (8.25,-2.1) [graysum] (diffsign) {+};
		\node at (8.25,-0.5) [graysum] (diffsign) {+};
		\node at (8.25,1.05) [graysum] (diffsign) {+};
		\node at (8.25,2.6) [graysum] (diffsign) {+};
		\node at (8.25,3.95) [graysum] (diffsign) {+};

		\draw[every loop,line width=2.3mm,>=latex,draw=myorange,fill=myorange]
		(-1.9,1.1) edge[] (-1.9+\x,1.1)
		(-0.5,0.15) edge[] (-0.5+\x,0.15)
		(1.1,-0.75) edge[] (1.1+\x,-0.75)
		(2.7,-1.7) edge[] (2.7+\x,-1.7)
		(4.6,-2.6) edge[] (4.6+\x,-2.6)
		(6.3,-3.5) to[] (6.3+\x,-3.5);
		
		\networkLayer{3.5}{0.05}{0.0}{0.0}{black!60,fill opacity=0.4}{\large$128{\times}128{\times}16$}{input}
		\networkLayer{3.0}{0.1}{2.4}{2.0}{black!60,fill opacity=0.4}{\large$64{\times}64 {\times}32$}{conv1}	
		\networkLayer{2.5}{0.2}{4.7}{4.0}{black!60,fill opacity=0.4}{\large$32{\times}32{\times}64$}{conv2}
		\networkLayer{2.0}{0.4}{6.9}{6.0}{black!60,fill opacity=0.4}{\large$16{\times}16{\times}128$}{conv3}   
		\networkLayer{1.5}{0.8}{9.1}{8.0}{black!60,fill opacity=0.4}{\large$8{\times}8{\times}256$}{conv4}   
		\networkLayer{1.0}{1.6}{11.4}{10.0}{black!60,fill opacity=0.4}{$4{\times}4 {\times}512$}{conv5}

		\node at (18.2,1.6) [] {\textbf{\textcolor{myturquoise}{\Large ${\circledast}({\uparrow}2)$}}};
		\draw[every loop,line width=2.3mm,>=latex,draw=myturquoise,fill=myturquoise]
		(17.5,1.0) edge[] (17.5+\x,1.0)
		(15.8,0.15) edge[] (15.8+\x,0.15)
		(14.0,-0.75) edge[] (14.0+\x,-0.75)
		(12.4,-1.7) edge[] (12.4+\x,-1.7)
		(10.8,-2.6) edge[] (10.8+\x,-2.6)
		(8.9,-3.5) to[] (8.9+\x,-3.5);	

		\networkLayer{1.5}{0.8}{13.5}{8.0}{black!60,fill opacity=0.4}{\large$~~8{\times}8{\times} 256$}{deconv4} 
		\networkLayer{2.0}{0.4}{14.7}{6.0}{black!60,fill opacity=0.4}{\large$~~~~~~16{\times}16{\times}128$}{deconv3}  
		\networkLayer{2.5}{0.2}{15.8}{4.0}{black!60,fill opacity=0.4}{\large$~~~~32{\times}32{\times}64$}{deconv2}
		\networkLayer{3.0}{0.1}{16.9}{2.0}{black!60,fill opacity=0.4}{\large$~~~64{\times}64{\times}32$}{deconv1}
		\networkLayer{3.5}{0.1}{18.1}{0.0}{black!60,fill opacity=0.4}{\large$~~128{\times}128{\times}16$}{deconv0}
		\node[canvas is zy plane at x = 0.2, transform shape,anchor=south,fill opacity=0.95] at (19.35,-0.0){\includegraphics[width=4cm,height=4.5cm]{imgs/introDiagram/clean.png}};

	\newcommand\mysize{2.5}
	\begin{scope}[xshift=3.4cm, yshift= -6.5 cm]
	\draw[every loop,line width=2.3mm,>=latex,draw=myorange,fill=myorange]
		(0,0) to[] (\x,0);
		\node (orange) at (1.8,0) [fontscale=\mysize,anchor=west] {\begin{varwidth}{8cm} 
		Strided convolution (5x5)\\
		+ BN + LeakyReLU
		\end{varwidth}};
		
	\draw[every loop,line width=2.3mm,>=latex,draw=myturquoise,fill=myturquoise]
		(0,-2.5) to[] (\x,-2.5);
		\node (orange) at (1.8,-2.5) [fontscale=\mysize,anchor=west] {\begin{varwidth}{8cm} 
		Upsampling (${\uparrow}2$) \\
		+ Convolution (5x5) \\
		+ BN + LeakyReLU
		\end{varwidth}};
	
	\newcommand\xshift{9.0}
	\node at (\xshift+0.6,0.7) [] {\textbf{\textcolor{myturquoise}{\Large ${\circledast}({\uparrow}2)$}}};
	\draw[every loop,line width=2.3mm,>=latex,draw=myturquoise,fill=myturquoise]
		(\xshift,0.0) to[] (\xshift+\x,0.0);
	\node (orange) at (\xshift+1.8,0.0) [fontscale=\mysize,anchor=west] {\begin{varwidth}{8cm} 
		Upsampling (${\uparrow}2$) \\
		+ Strided convolution (5x5) \\
		+ Sigmoid
		\end{varwidth}};
		
	\draw [->,color=black!60,line width=0.9mm] 
		(\xshift,-2.5) to [bend left=\bending] (\xshift+\x,-2.5);
	\node at (\xshift+0.635,-2.4) [graysum] (diffsign) {+};
	\node (blue) at (\xshift+1.8,-2.5) [fontscale=\mysize,anchor=west] {\begin{varwidth}{6cm} 
		Addition (only for AESc)
		\end{varwidth}};
	\end{scope}
\end{tikzpicture}

%% file: imgs/StainModel/ellipse.tex
\begin{tikzpicture}[node distance=2cm,>=latex', every text node part/.style={align=center}]

\begin{scope}[xshift=-20cm] 
\begin{axis}[scale only axis=true, axis line style={draw=none}, xtick=\empty,ytick=\empty]
	\addplot[ultra thick,myblue!30,smooth,tension=0.75,dashed] table[col sep=comma] {imgs/StainModel/initial_ellipse.dat};
	\addplot[only marks,myblue,fill=myblue!50, mark size=3pt] table[col sep=comma] {imgs/StainModel/initial_ellipse.dat};
	
	\addplot[only marks,color=mycolor,fill=mycolor!70, mark size=3pt] table[col sep=comma] {imgs/StainModel/perturbated_ellipse.dat};
	\addplot[ultra thick,mycolor!70,fill=mycolor,fill opacity=0.2] table[col sep=comma] {imgs/StainModel/interpolation.dat};
	
	\addplot[color=mycolor!70] coordinates {(607,793) (542,825)};
	\addplot[color=mycolor!70] coordinates {(723,769) (782,755)};
	\addplot[color=mycolor!70] coordinates {(839,721) (824,717)};
	\addplot[color=mycolor!70] coordinates {(952,628) (894,564)};
	\addplot[color=mycolor!70] coordinates {(1000,512) (1051,577)};
	\addplot[color=mycolor!70] coordinates {(1000,492) (903,515)};
	\addplot[color=mycolor!70] coordinates {(956,376) (1028,461)};
	\addplot[color=mycolor!70] coordinates {(844,282) (822,228)};
	\addplot[color=mycolor!70] coordinates {(728,232) (745,231)};
	\addplot[color=mycolor!70] coordinates {(612,207) (673,140)};
	\addplot[color=mycolor!70] coordinates {(500,199) (559,179)};
	\addplot[color=mycolor!70] coordinates {(384,207) (475,138)};
	\addplot[color=mycolor!70] coordinates {(268,233) (333,212)};
	\addplot[color=mycolor!70] coordinates {(152,284) (234,259)};
	\addplot[color=mycolor!70] coordinates {(41,380) (24,282)};
	\addplot[color=mycolor!70] coordinates {(0,496) (69,414)};
	\addplot[color=mycolor!70] coordinates {(1,516) (74,421)};
	\addplot[color=mycolor!70] coordinates {(51,632) (8,698)};
	\addplot[color=mycolor!70] coordinates {(165,722) (69,726)};
	\addplot[color=mycolor!70] coordinates {(281,770) (292,678)};
	\addplot[color=mycolor!70] coordinates {(397,794) (341,856)};
	
\end{axis}
\end{scope}

\begin{scope}[xshift=-9.5cm,yshift=2.5cm] 
\draw[->,color=black!60,line width=1.5mm] (-2.0,1.0) to [bend left=20] (2.0,1.0);
\node at (0,2.7) (input)[text width=5cm]{\huge Add the irregular structure to the image};
\node at (0,-1.0) (input)[text width=5cm]{\huge Randomize the size, the color, the location and the general shape};
\end{scope}

\begin{scope}[xshift=2cm] 
\node at (-6.0,3.3) (input) {\includegraphics[width=5.5cm]{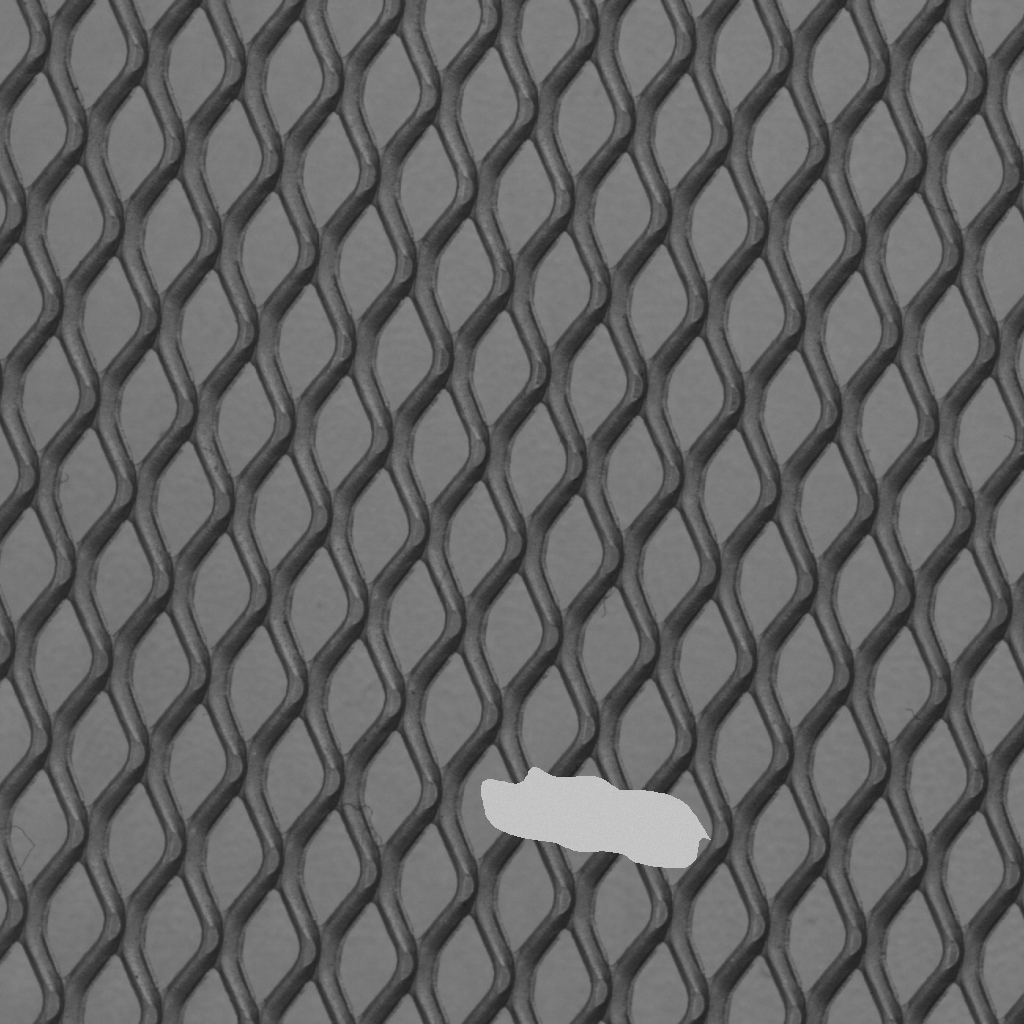}};
\node at (0,3.3) (input) {\includegraphics[width=5.5cm]{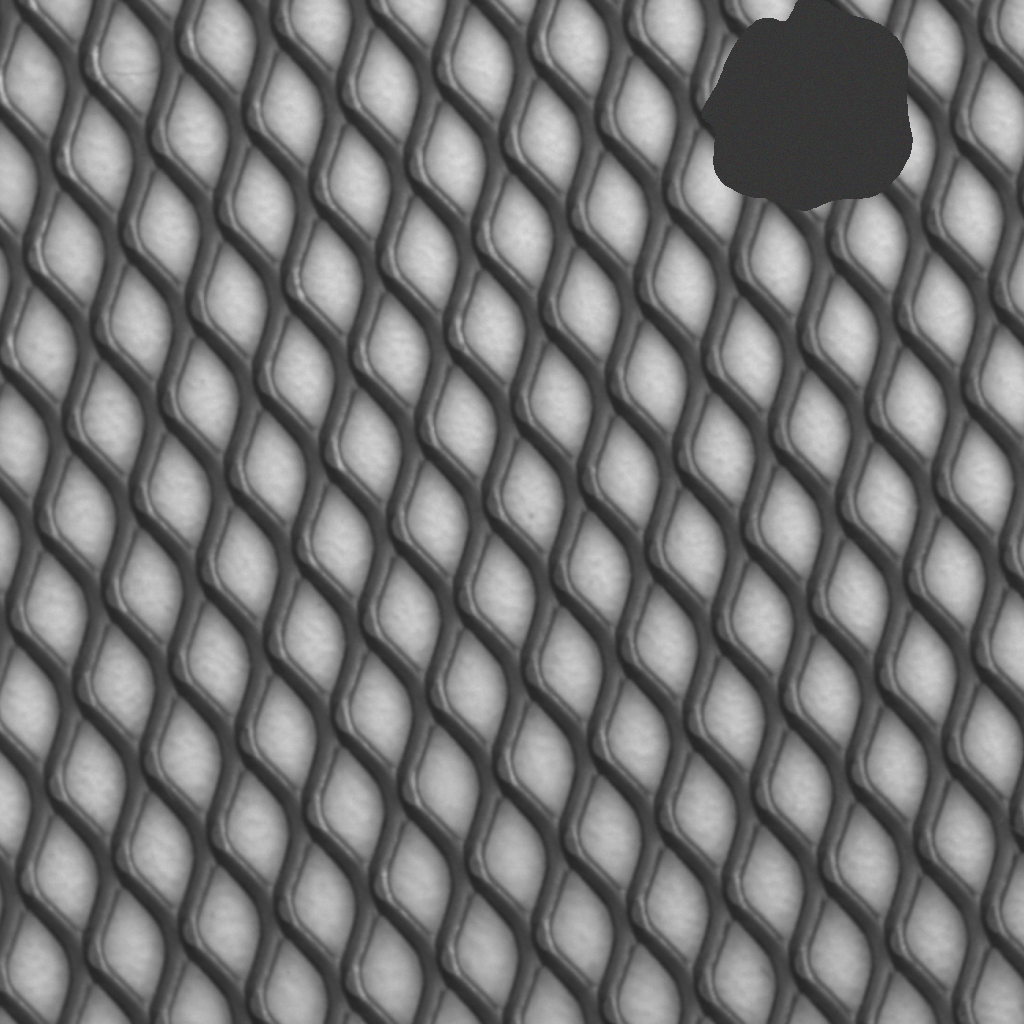}};
\node at (6.0,3.3) (input) {\includegraphics[width=5.5cm]{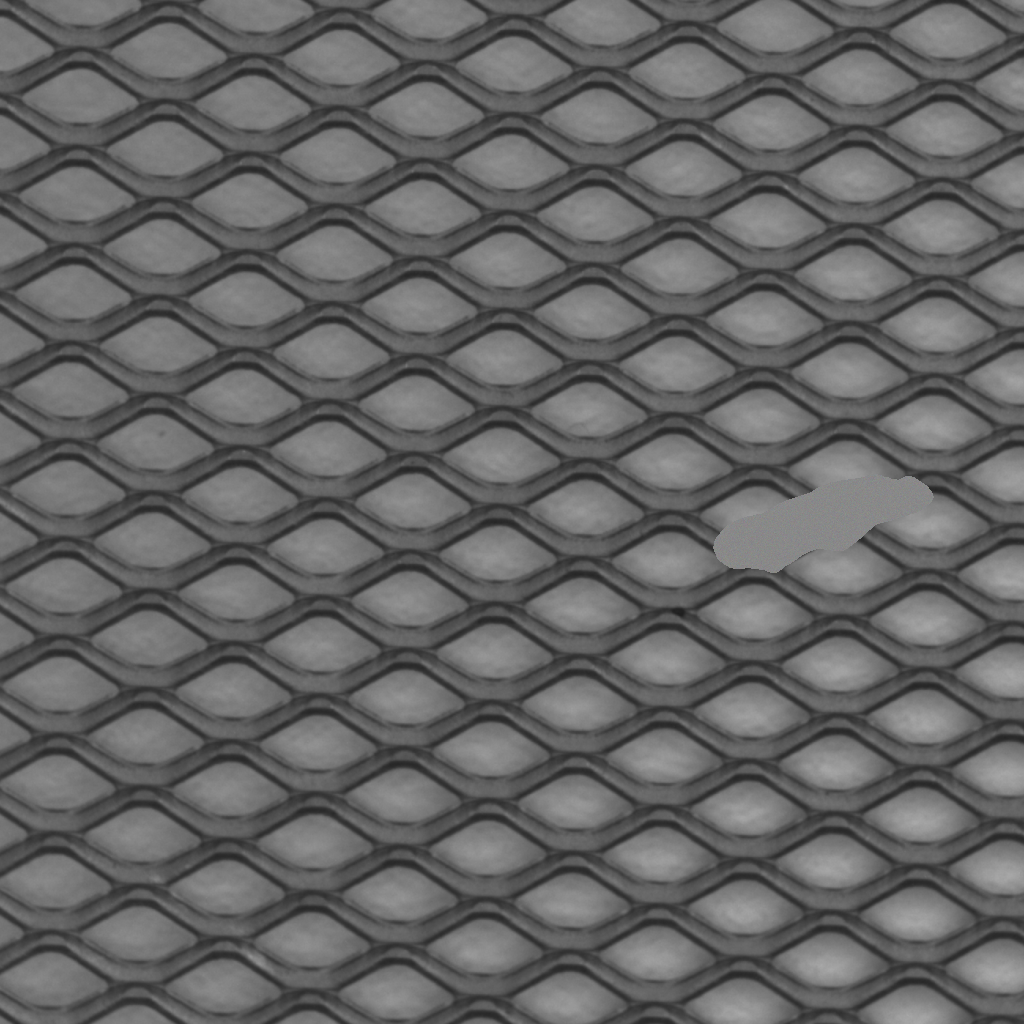}};
\end{scope}
  
\end{tikzpicture}

%% file: Chapters/Results.tex
Experiments have been conducted on grayscale images of the MVTec AD dataset  \cite{Bergmann2019}, containing 5 categories of textures and 10 categories of objects. In this dataset, defects are real and have various appearance. Their location is defined with a binary segmentation mask. All images have been scaled to a $256\times256$ size. Anomaly detection is performed at this resolution.


\subsection{AESc + Stain: qualitative and quantitative analysis} \label{sec:resultanalysis}

In this section, we compare qualitatively and quantitatively the results obtained with our AESc + Stain model for both image- and pixel-wise detection. We focus this first analysis on the residual-based detection approach to emphasize the benefits of adding skip connections to the AE architecture. The comparison of the results obtained with  residual- versus uncertainty-based strategies is discussed later in Section \ref{sec:residualvsuncertainty}. \\
\begin{figure}[!t]
\centering
\input{imgs/Results/smallComparison_CarpetReal19}
\input{imgs/Results/smallComparison_GridReal15}
\input{imgs/Results/smallComparison_CableReal61}
\input{imgs/Results/smallComparison_ToothbrushReal5}
\caption{Predictions obtained with the AESc and AE networks trained with and without our Stain noise model. Two defective textures are considered, namely a carpet (first sample) and a grid (second sample), as well as two defective objects, namely a cable (third sample) and a toothbrush (fourth sample). First column show the image fed in the networks and the mask locating the defect. Odd rows show the reconstructed images and even rows show the anomaly maps obtained with the residual-based strategy.}
\label{fig:qualitativeRes}
\end{figure}
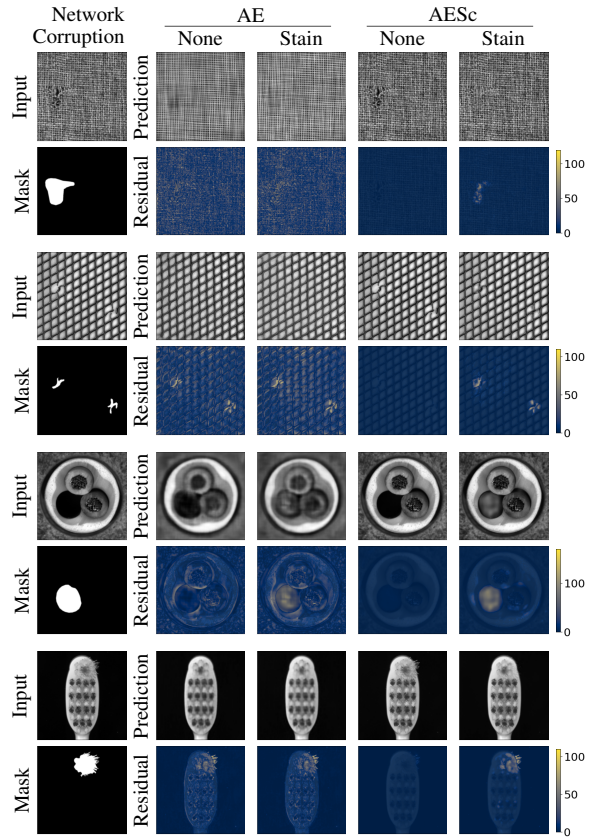

\input{Tables/RESULT_image-wise}

\input{Tables/RESULT_pixel-wise}

Qualitatively, Figure \ref{fig:qualitativeRes} reveals the general trends of the AE and AESc models trained with and without the Stain noise corruption. On the one hand, the AE network produces blurry reconstructions as depicted by the overall higher residual intensities. If the global structure of the object images (cable and toothbrush) are properly reconstructed, the AE network struggles to infer the finer details of the texture images (carpet sample). On the other hand, the AESc model shows finer reconstruction of the image details depicted by a nearly zero residual over the clean areas of the images. However, when ASEc is trained without corruption, the model converges towards an identity operator, as revealed by the close-to-zero residuals of defective structures. The corruption of the training images with the Stain model alleviates this unwanted behavior by leading to high reconstruction residuals in defective areas while simultaneously keeping low reconstruction residuals in clean structures.\\

Quantitatively, the image-wise detection performances obtained with the AESc and AE networks trained with and without our Stain noise model are presented in Table \ref{table:summary_image-wise}. The last column provides a comparison with the ITEA method, introduced by Huang et al. \cite{Huang2019}. ITAE is also a reconstruction-based approach which relies on an autoencoder with skip connections trained with images corrupted by random rotations and a graying operator (averaging of pixel value along the channel dimension) selected based on prior knowledge about the task. \\
This table highlights the superiority of our AESc + Stain noise model to solve the image-wise anomaly detection. The improvement brought by adding skip connections to an autoencoder trained with corrupted images is even more important for texture images than for object images. We also observe that, if the highest accuracy is consistently obtained with the residual-based approach, the uncertainty-based decision derived from the AESc + Stain model generally provides the second best (underlined in Table \ref{table:summary_image-wise}) performances among tested networks, attesting the quality of the AESc + Stain model for image-based decision.\\

Table \ref{table:summary_pixel-wise} presents the pixel-wise detection performances obtained with our approaches and compares them with the method reported in \cite{Bergmann2019}, referred to as AE\textsubscript{L2}. This residual-based method relies on an autoencoder without skip connections, and provides SoA performance in the pixel-wise detection scenario. Similarly to our AE model, AE\textsubscript{L2} is trained to minimize the MSE of the reconstruction of images that are not corrupted with synthetic noise. AE\textsubscript{L2} however differs from our AE model in several aspects, including a different network architecture, data augmentation, patch-based inference for the texture images, and anomaly map post-processing with mathematical morphology. Despite our efforts, in absence of public code, we have been unable to reproduce the results presented in \cite{Bergmann2019}. Hence, our table just copy the results from \cite{Bergmann2019}. For fair comparison between AE and AESc + Stain, the table also provides the results obtained with our AE, since our AE and AESc + Stain models adopt the same architecture (up to the skip connections) and the same training procedure.\\
In the residual-based detection strategy, our AESc + Stain method obtains similar performances as the AE\textsubscript{L2} approach when averaged over all the image categories of the MVTec AD dataset. However, as already pointed in the image-wise detection scenario, we notice that AESc + Stain performs better with texture images and worse with object images. Regarding the decision strategy, we observe an opposite trend than the one encountered for image-wise detection: the uncertainty-based approach performs a bit better than the residual-based strategy when it comes to pixel-wise decisions. This difference is further investigated in the next section.\\



\subsection{Residual- vs. uncertainty-based detection strategies} \label{sec:residualvsuncertainty}

\begin{figure}[!t]
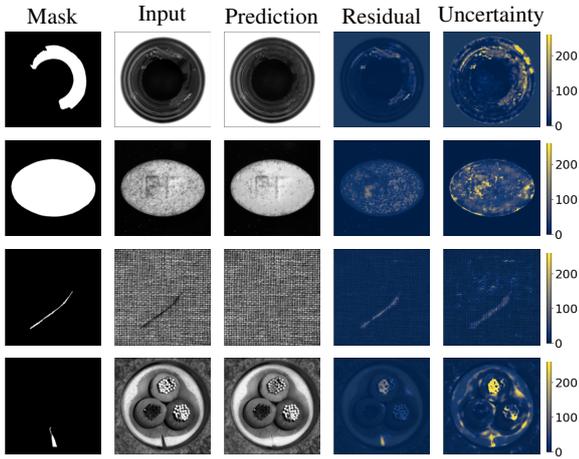

\hspace*{-2.25cm}
\includestandalone[width=11.0cm]{imgs/Results/residualVSuncertainty}
\vspace*{-0.4cm}
\caption{Predictions obtained with the AESc network trained with our Stain noise model. One defective texture is considered, namely a carpet (third row) as well as three defective objects, namely a bottle (first row), a pill (second row) and a cable (fourth row). From left to right, columns represent the ground-truth, the image fed to the network, the prediction (without MCDropout), the reconstruction residual and the reconstruction uncertainty.
}
\label{fig:residualVSuncertainty}
\end{figure}

Figure \ref{fig:residualVSuncertainty} provides a visual comparison between residual- and uncertainty-based strategies. Globally, we observe that the reconstruction residual mostly correlates with the uncertainty. However, the uncertainty indicator is usually more widespread. This behavior can sometimes lead to a better coverage of the defective structures (bottle and pill) or increase the number of false positive pixels that are detected (carpet and cable). \\
One important observation concerns the relationship between the detection of a defective structure and its contrast with its surroundings. In the residual-based approach, regions of an image are considered as defective if their reconstruction error exceeds a threshold. In the proposed formulation, the network is explicitly constrained to replace synthetic defective structures with clean content. No constraint is introduced regarding the contrast of the reconstructed structure and its surroundings. Hence, defects that are poorly contrasted lead to small residual intensities. On the contrary, the intensity of the uncertainty indicator does not depend on the contrast between a structure with the surroundings. For low contrast defects, it enhances their detection as illustrated (bottle and pill). On the contrary, it can deteriorate the location of high contrast defects for which the residual map is an appropriate anomaly indicator (carpet and cable). In theses cases, the sharp prediction obtained with the residual-based approach is preferred over the uncertainty-based one. \\

As reported in Section \ref{sec:resultanalysis}, we observe that the uncertainty-based detection perform generally worse than the residual-based approach for image-wise detection. We explain this drop of performance by an increase of the intensities of the uncertainty maps inferred from the clean images belonging to the test set. As the image-wise detection is based on the $\mathcal{L}^2$ norm of the anomaly map, the lowest the anomaly maps of clean images, the better the detection of defective images. For image-wise detection, the performances are less sensitive to the optimal coverage of the defective area as long as the overall intensity of the clean anomaly maps is low.\\
On the contrary, the uncertainty-based strategy improves the pixel-wise detection of the AESc + Stain model. For this use case, a better coverage of the defective structure is crucial. As previously mentioned, AESc + Stain model used usually leads to reconstruction residual constituted of sporadic spots and misses low contrast defects. The uncertainty-based strategy compensates these two issues.


\subsection{Comparative study of corruption models} \label{sec:residualvsuncertainty}

\begin{figure}[!ht]
\centering
      \subfloat[Gaussian noise.]{\quad\includegraphics[width=1.8cm]{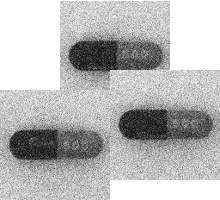}\quad}
      \subfloat[Scratch.]{\quad\includegraphics[width=1.8cm]{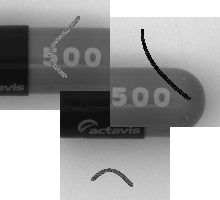}\quad}
      \subfloat[Drops.]{\quad\includegraphics[width=1.8cm]{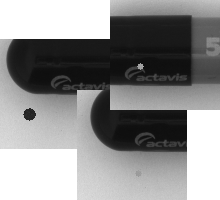}\quad}
\caption{Illustration of the Gaussian noise, Scratch and Drops models. The original clean image is the one presented in Figure \ref{fig:introDiagram}.}
\label{fig:newCorruption}
\end{figure}

Up to now, we considered only the Stain noise model to corrupt training data. In this comparative study we consider other noise models to confirm the relevance of our previous approach over other types of corruption that could have been considered. We provide here a comparison with three other synthetic noise models represented in Figure \ref{fig:newCorruption}:
\begin{description}
\item[a- Gaussian noise.] Corrupt by adding white noise applied uniformly over the entire image. For normalized intensities between 0 and 1, a corrupted pixel value $x'$, corresponding to an initial pixel value $x$, is the realization of a random variable given by a normal distribution of mean $x$ and variance $\sigma^2$ in the set: $[0.1, 0.2, 0.4, 0.8]$.
\item[b- Scratch.] Corrupt by adding one curve connecting two points whose coordinates are randomly chosen in the image and whose color is randomly selected in the gray scale range. The curve can follow a straight line, a sinusoidal wave or the path of a square root function. 
\item[c- Drops.] Corrupt by adding 10 droplets whose color are randomly selected in the gray scale range and whose shape are circular with a random diameter (chosen between 1 and 2\% of the smallest image dimension). The droplets partially overlap. 
\end{description}

\vspace*{0.5cm}

In addition, we have also considered the possibility to corrupt the training images with a combination of several models. We propose two hybrid models:
\begin{description}
\item[d- Mix1.] This configuration corrupts training images with a combination of the Stain, Scratch and Drops models. We fix that 60\% of the training images are corrupted with the Stain model while the remaining 40\% are corrupted with the Scratch and Drops models in equal proportions. 
\item[e- Mix2.] This configuration corrupts training images with a combination of the Stain and the Gaussian noise models. We fix that 60\% of the training images are corrupted with the Stain model while the remaining 40\% are corrupted with the Gaussian noise model.
\end{description}

\begin{figure*}[!t]
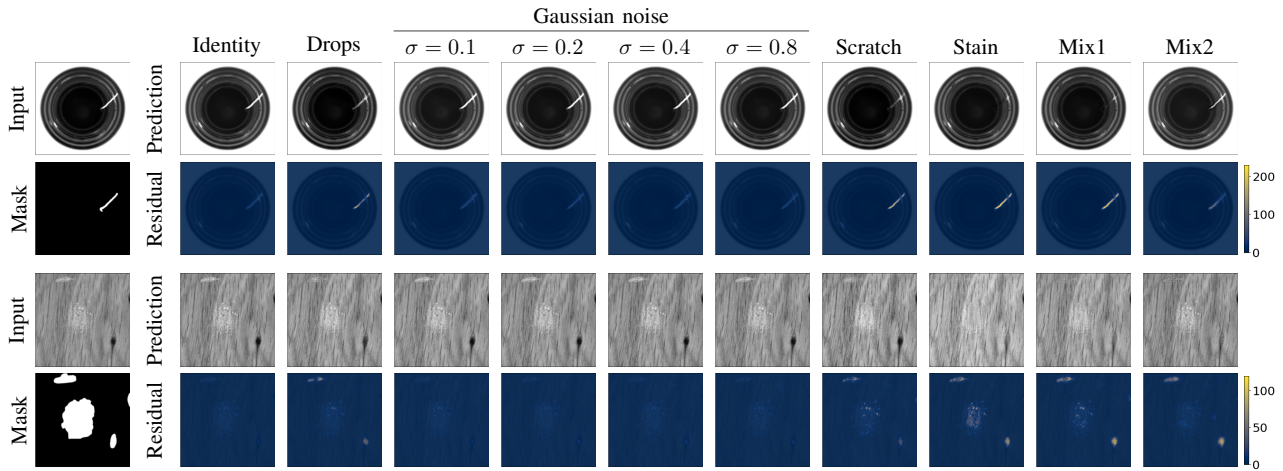

\hspace*{-5.5cm}\includestandalone[width=1.32\textwidth]{imgs/Results/corruptionComparison_BottleReal60}
\hspace*{-5.5cm}\includestandalone[width=1.32\textwidth]{imgs/Results/corruptionComparison_WoodReal8}
\vspace*{-0.4cm}
\caption{Reconstructions obtained with the AESc network trained with different noise models. We consider here one defective object, namely a bottle (first sample) and a defective texture, namely a wood (third sample). Rows and columns are defined as in Figure \ref{fig:qualitativeRes}.} 
\label{fig:aesc_corruptionComparison}
\end{figure*}

Figure \ref{fig:aesc_corruptionComparison} allows the comparison of the newly introduced noise models with the Stain one over similar samples. First, these examples illustrate the convergence of the model towards the identity mapping when the Gaussian noise model is used as synthetic corruption. An analysis of the results obtained over the entire dataset reveals that the AESc + Gaussian noise model does almost not differ from the AESc network trained with unaltered images. \\
Compared to the the Gaussian noise, other models introduced before improve the identification of defective areas in the images. This is reflected by higher intensities of the reconstruction residual in the defective areas and close-to-zero reconstruction residual in the clean areas. With the exception of the Gaussian noise model, the Scratch model is the most conservative, among those considered, in the sense that most of the structures of the input images tend to be reconstructed identically. This practice increases the number of false negative. Also, the Drops model restricts the structures detected as defective to sporadic spots. Finally, the three models based on the Stain noise (Stain, Mix1 and Mix2) provide the residuals that correlate the most with the segmentation mask. \\
Generally, models based on the Stain noise (Stain, Mix1 and Mix2) lead to the most relevant reconstruction for anomaly detection, i.e. lower residual intensities in clean areas and higher residual intensities in defective areas. More surprisingly, this statement remains true even if the actual defect looks more similar to the Scratch model than the Stain noise (bottle sample in Figure \ref{fig:aesc_corruptionComparison}). We recall that defects contained in the MVTec AD dataset are real observations of an anomaly. This reflects that models trained with synthetic corruption models that look similar to real ones do not necessarily generalize well to real defects \\

\input{Tables/APPENDIX_image-wise}

Table \ref{table:AESc-residual-image} quantifies the impact of the synthetic noise model on the performances of the ASEc network to solve the image-wise detection task with a residual-based approach. The AESc + Stain configuration is the best performing in all use cases when considering the mean performances that are obtained over the entire dataset, as revealed by the previous qualitative study. The two hybrid models (Mix1 and Mix2) lead usually to sightly lower performances than those obtained with the Stain model. Those observations attest that the Stain model is superior to others and justify the choice of the Stain noise as our newly introduced approach to corrupt the training images with synthetic noise.

%% file: imgs/Results/smallComparison_CarpetReal19.tex
{ 

\newif\ifMC
    \MCfalse  
    
\newif\iflegend
    \legendtrue   
    
\newcommand\inputPATH{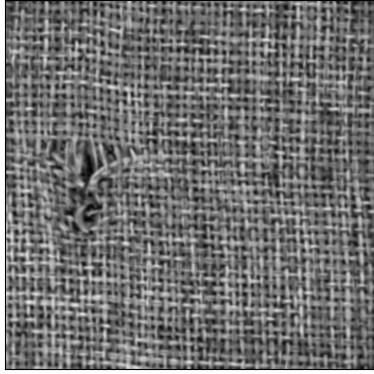}
\newcommand\maskPATH{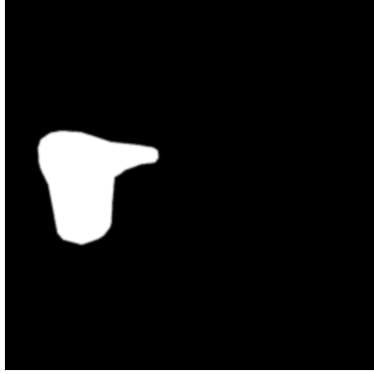}
\newcommand\colorbarResidualPATH{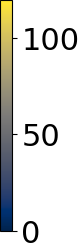}
\newcommand\colorbarUncertaintyPATH{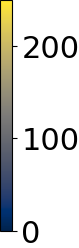}

\newcommand\CAEpredPATH{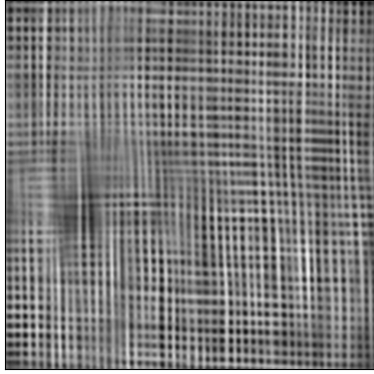}
\newcommand\CAEresidualPATH{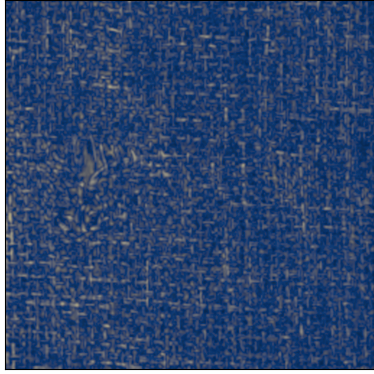}
\newcommand\CAEuncertaintyPATH{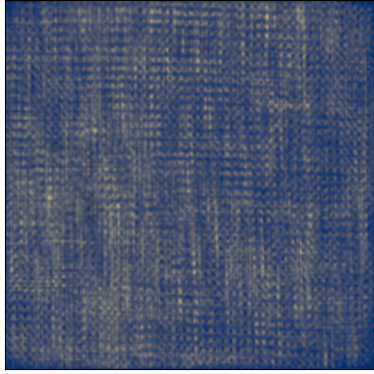}

\newcommand\stainCAEpredPATH{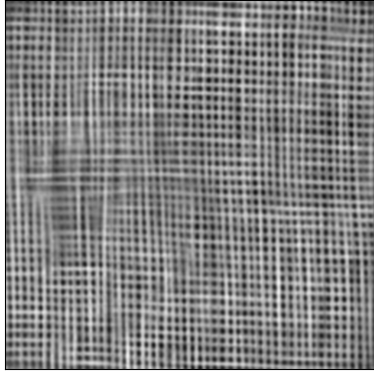}
\newcommand\stainCAEresidualPATH{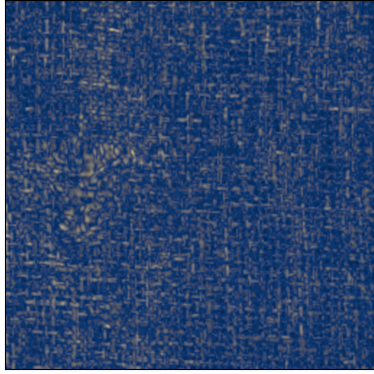}
\newcommand\stainCAEuncertaintyPATH{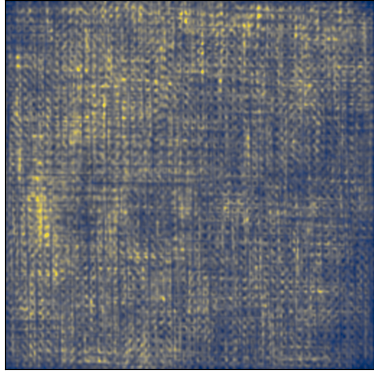}

\newcommand\CAESSCpredPATH{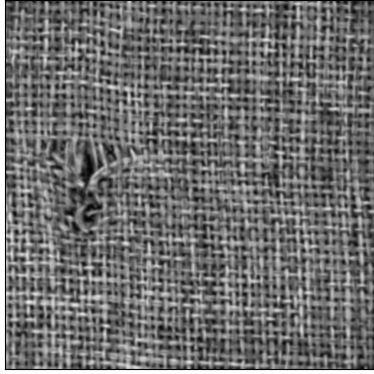}
\newcommand\CAESSCresidualPATH{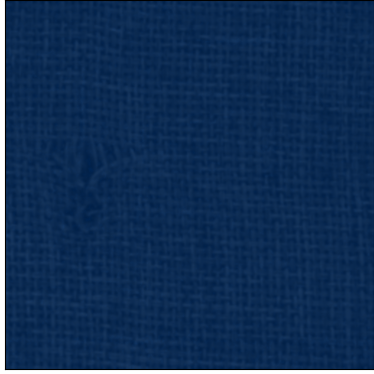}
\newcommand\CAESSCuncertaintyPATH{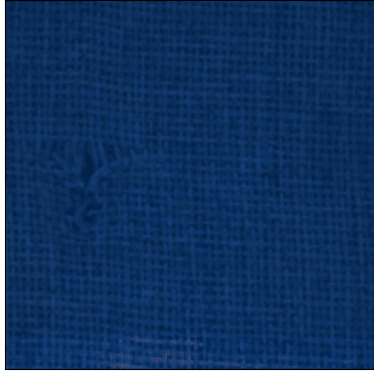}

\newcommand\stainCAESSCpredPATH{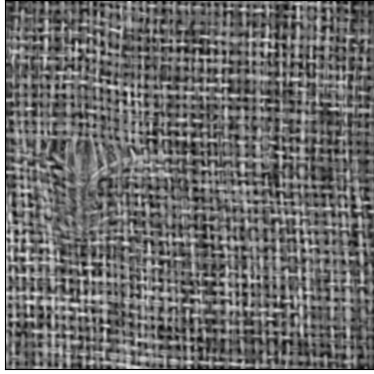}
\newcommand\stainCAESSCresidualPATH{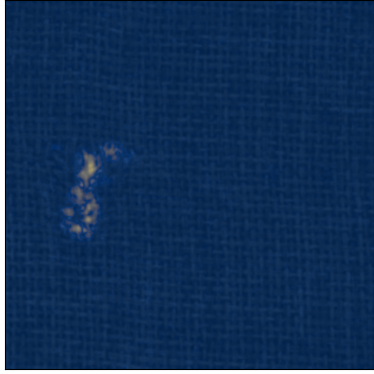}
\newcommand\stainCAESSCuncertaintyPATH{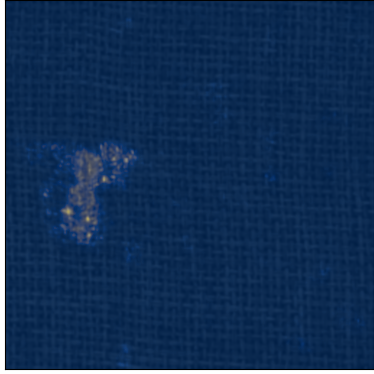}

\includestandalone[width=8.2cm]{imgs/Results/smallComparison}

}

%% file: imgs/Results/smallComparison_GridReal15.tex
{ 

\newif\ifMC
    \MCfalse  
    
\newif\iflegend
    \legendfalse 
    
\newcommand\inputPATH{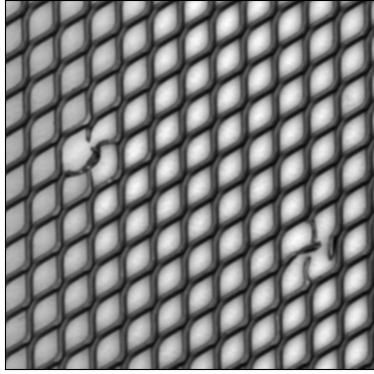}
\newcommand\maskPATH{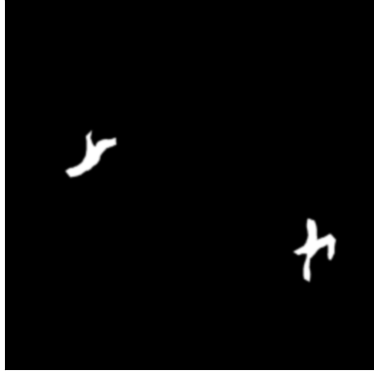}
\newcommand\colorbarResidualPATH{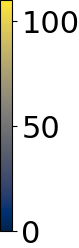}
\newcommand\colorbarUncertaintyPATH{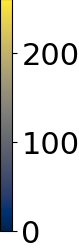}

\newcommand\CAEpredPATH{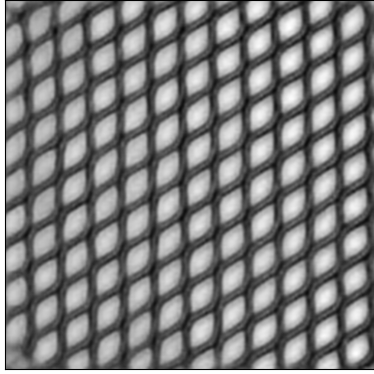}
\newcommand\CAEresidualPATH{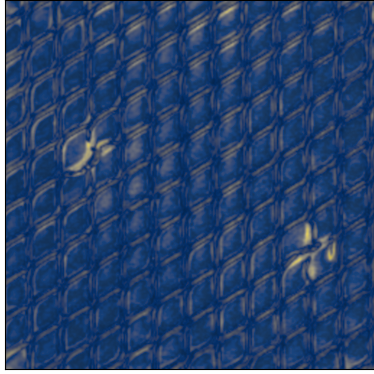}
\newcommand\CAEuncertaintyPATH{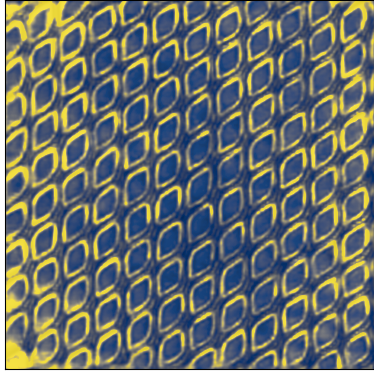}

\newcommand\stainCAEpredPATH{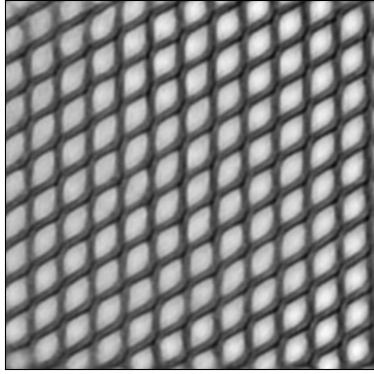}
\newcommand\stainCAEresidualPATH{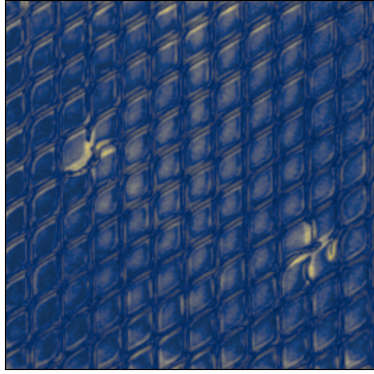}
\newcommand\stainCAEuncertaintyPATH{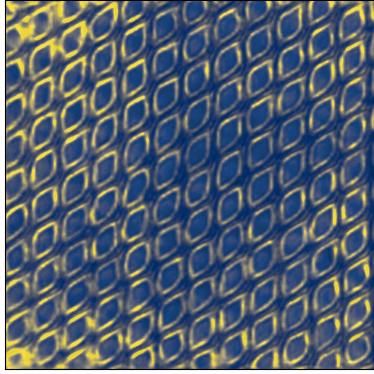}

\newcommand\CAESSCpredPATH{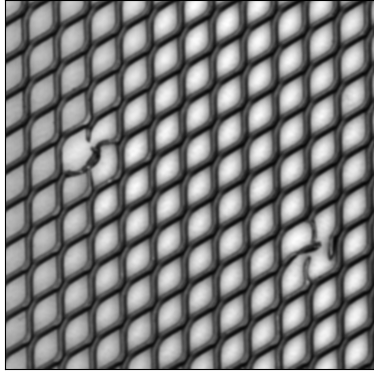}
\newcommand\CAESSCresidualPATH{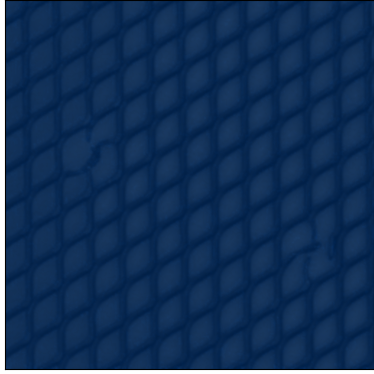}
\newcommand\CAESSCuncertaintyPATH{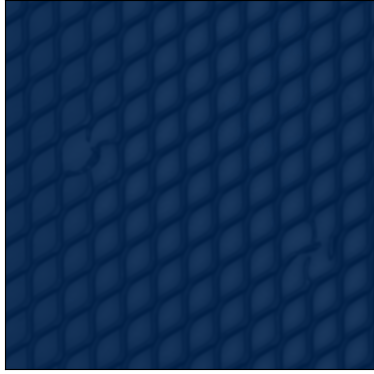}

\newcommand\stainCAESSCpredPATH{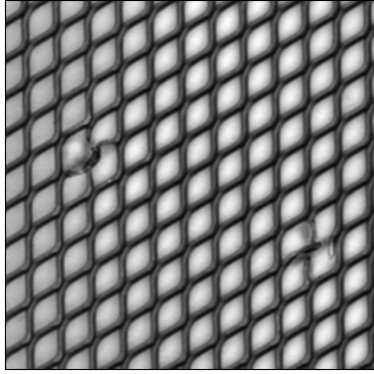}
\newcommand\stainCAESSCresidualPATH{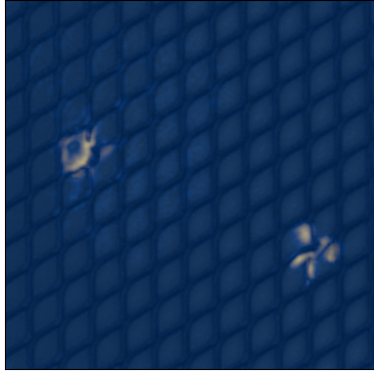}
\newcommand\stainCAESSCuncertaintyPATH{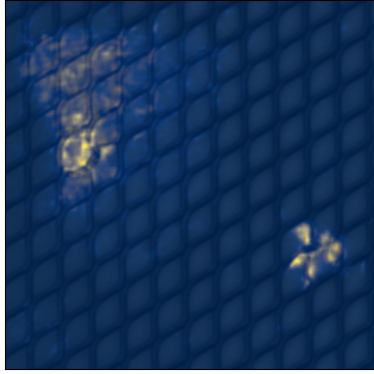}

\includestandalone[width=8.2cm]{imgs/Results/smallComparison}

}

%% file: imgs/Results/smallComparison_CableReal61.tex
{ 

\newif\ifMC
    \MCfalse  
    
\newif\iflegend
    \legendfalse 
    
\newcommand\inputPATH{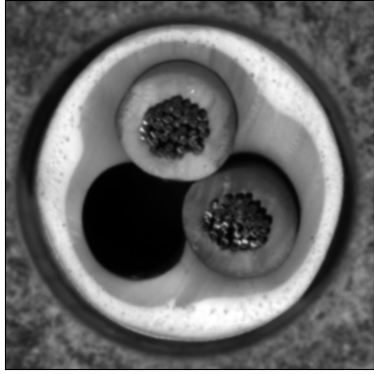}
\newcommand\maskPATH{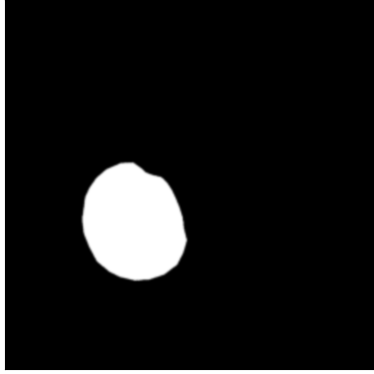}
\newcommand\colorbarResidualPATH{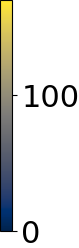}
\newcommand\colorbarUncertaintyPATH{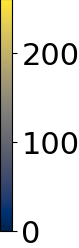}

\newcommand\CAEpredPATH{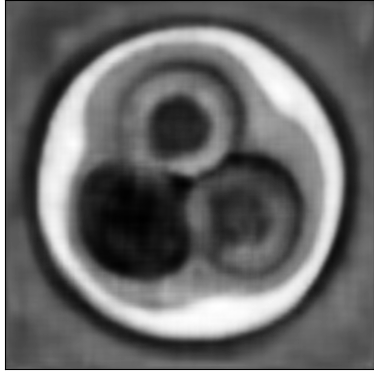}
\newcommand\CAEresidualPATH{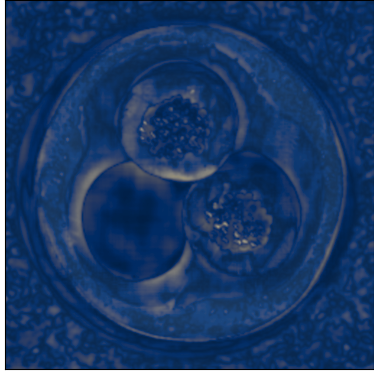}
\newcommand\CAEuncertaintyPATH{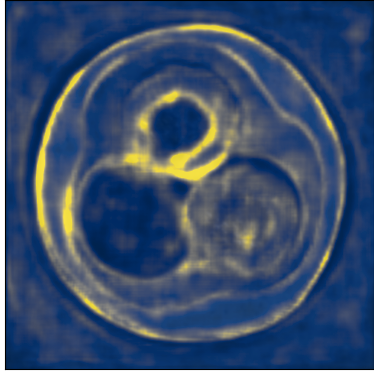}

\newcommand\stainCAEpredPATH{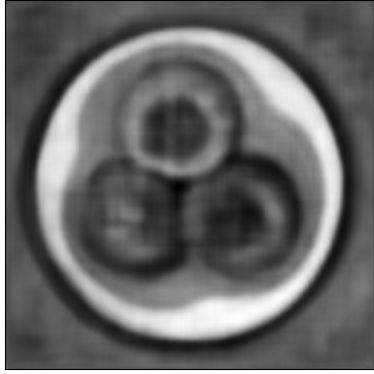}
\newcommand\stainCAEresidualPATH{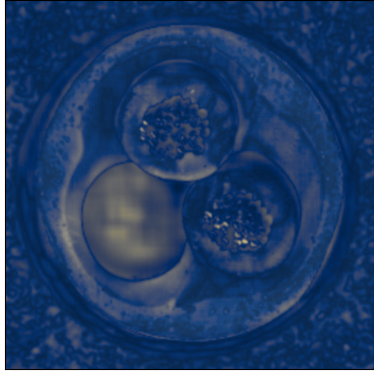}
\newcommand\stainCAEuncertaintyPATH{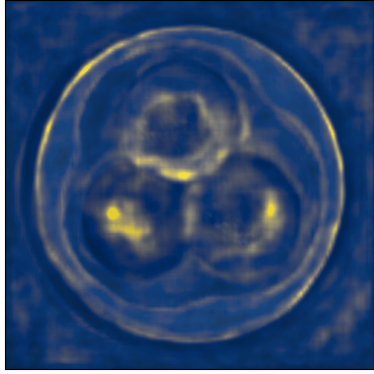}

\newcommand\CAESSCpredPATH{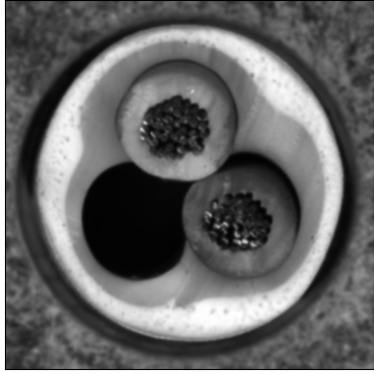}
\newcommand\CAESSCresidualPATH{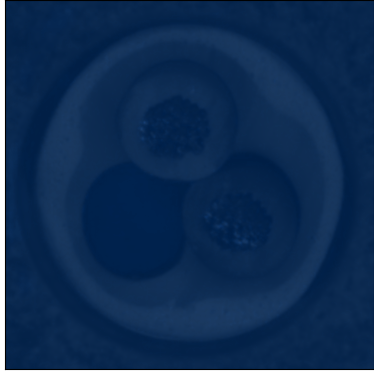}
\newcommand\CAESSCuncertaintyPATH{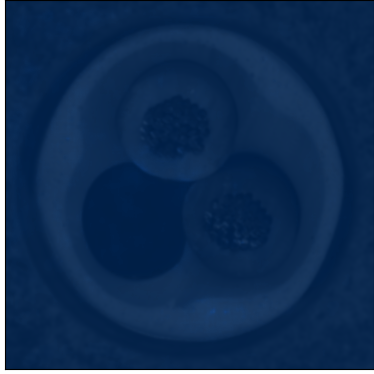}

\newcommand\stainCAESSCpredPATH{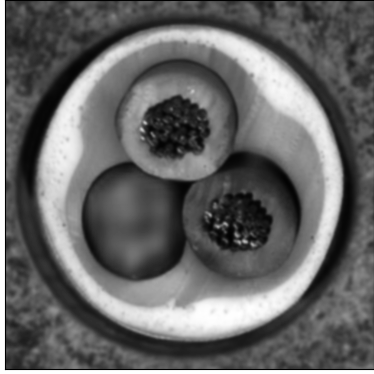}
\newcommand\stainCAESSCresidualPATH{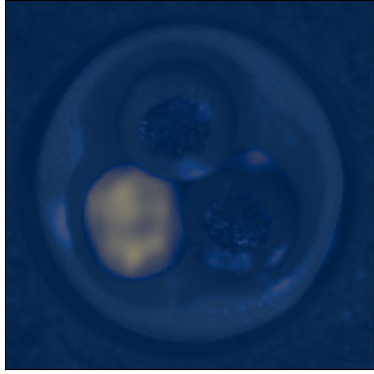}
\newcommand\stainCAESSCuncertaintyPATH{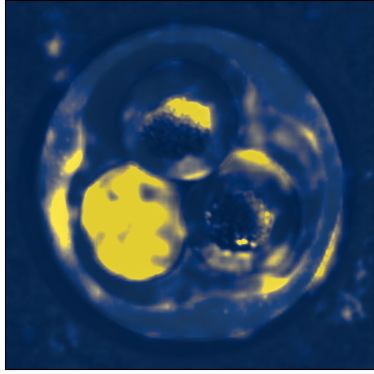}

\includestandalone[width=8.2cm]{imgs/Results/smallComparison}

}

%% file: imgs/Results/smallComparison_ToothbrushReal5.tex
{ 

\newif\ifMC
    \MCfalse  
    
\newif\iflegend
    \legendfalse 
    
\newcommand\inputPATH{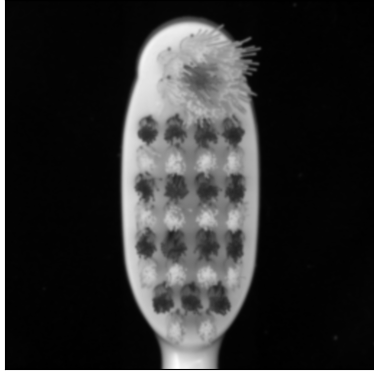}
\newcommand\maskPATH{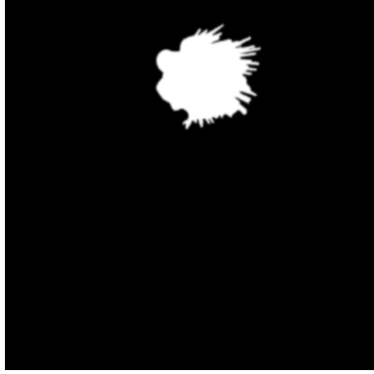}
\newcommand\colorbarResidualPATH{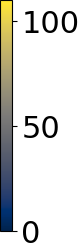}
\newcommand\colorbarUncertaintyPATH{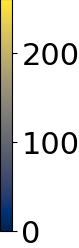}

\newcommand\CAEpredPATH{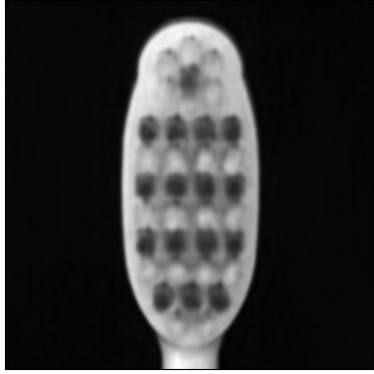}
\newcommand\CAEresidualPATH{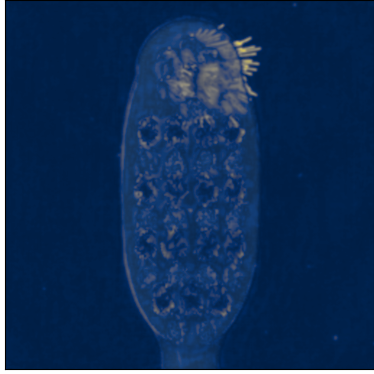}
\newcommand\CAEuncertaintyPATH{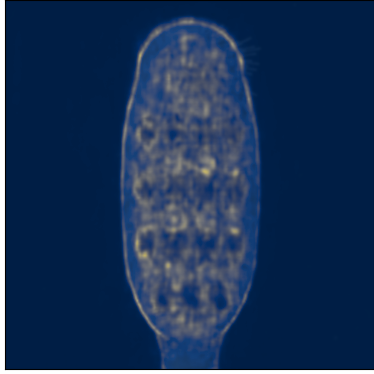}

\newcommand\stainCAEpredPATH{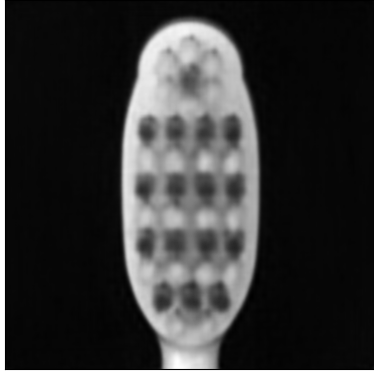}
\newcommand\stainCAEresidualPATH{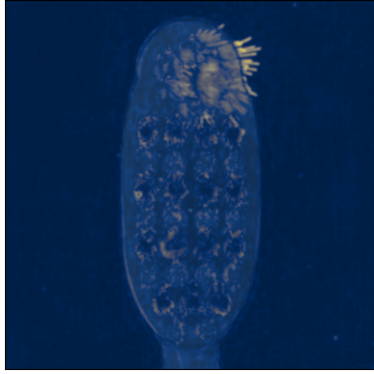}
\newcommand\stainCAEuncertaintyPATH{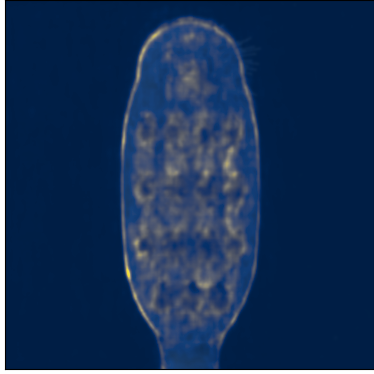}

\newcommand\CAESSCpredPATH{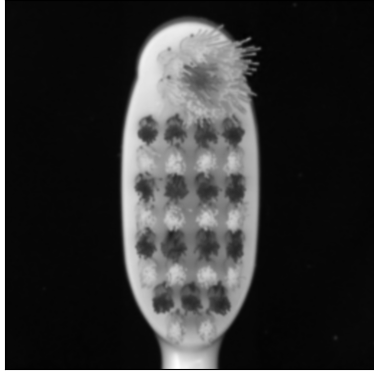}
\newcommand\CAESSCresidualPATH{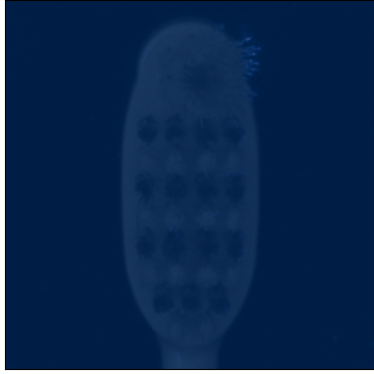}
\newcommand\CAESSCuncertaintyPATH{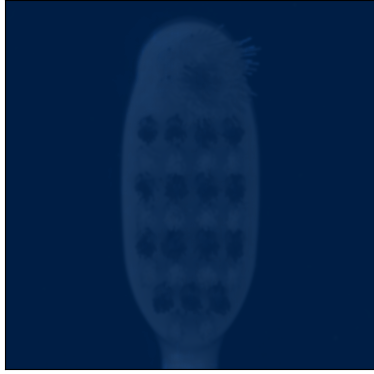}

\newcommand\stainCAESSCpredPATH{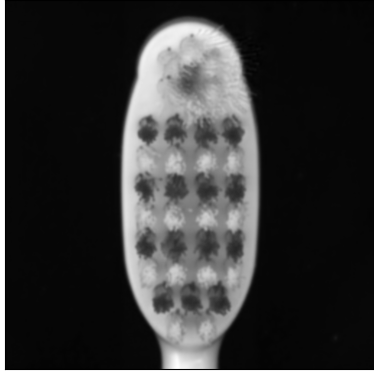}
\newcommand\stainCAESSCresidualPATH{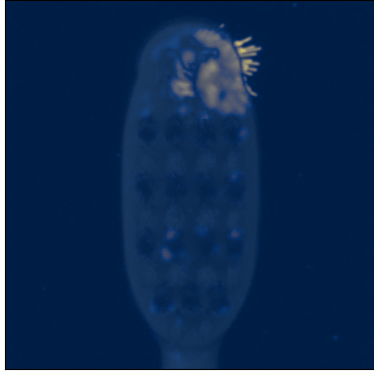}
\newcommand\stainCAESSCuncertaintyPATH{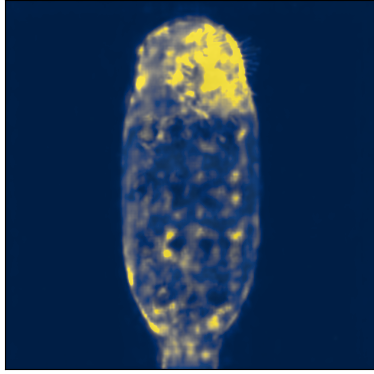}

\includestandalone[width=8.2cm]{imgs/Results/smallComparison}

}

%% file: Tables/RESULT_image-wise.tex
\tabcolsep=0.075cm
\begin{table}[!t]
\begin{threeparttable}[b]
\caption{Image-wise detection AUC obtained with the residual- and uncertainty-based detection methods\tnote{$a$}.  }
\label{table:summary_image-wise}
\scriptsize

\newcommand\len{0.55}

\begin{tabularx}{8.9cm}{@{}llp{\len cm}p{\len cm}p{\len cm}p{\len cm}p{0.15cm}p{\len cm}p{\len cm}p{\len cm}p{\len cm}p{1.2cm}@{}} \toprule

&& \multicolumn{4}{c}{\textbf{Uncertainty}} && \multicolumn{5}{c}{\textbf{Residual}}  \\
\arrayrulecolor{black!60}\cmidrule(rr){3-6}\cmidrule(rr){8-12}
\multicolumn{2}{l}{Network} & \multicolumn{2}{c}{\textbf{AE}} & \multicolumn{2}{c}{\textbf{AESc}} && \multicolumn{2}{c}{\textbf{AE}} & \multicolumn{2}{c}{\textbf{AESc}} & \textbf{ITAE \cite{Huang2019}} \\
\arrayrulecolor{black!60}\cmidrule(rr){3-4}\cmidrule(rr){5-6}\cmidrule(rr){8-9}\cmidrule(rr){10-11}
\multicolumn{2}{l}{Corruption} & None & Stain & None & Stain && None & Stain & None & Stain &  \\

\arrayrulecolor{black!60}\midrule\midrule

\multirow{5}{*}{\rotatebox[origin=c]{90}{Textures}} & Carpet  & 0.41 & 0.30 & 0.44 & \underline{0.80} && 0.43 & 0.43 & 0.48 & \textbf{0.89} & 0.71 \\
& Grid &0.69 & 0.66 & 0.12 & \textbf{0.97} && 0.80 & 0.84 & 0.52 & \textbf{0.97} & \underline{0.88} \\
& Leather & 0.86 & 0.57 & \underline{0.88} & 0.72 && 0.45 & 0.54 & 0.56 & \textbf{0.89} & 0.87\\
& Tile & 0.73 & 0.50 & 0.72 & \underline{0.95} && 0.49 & 0.57 & 0.88 & \textbf{0.99} & 0.74\\
& Wood & 0.87 & 0.86 & 0.78 & 0.78 && 0.92 & \underline{0.94} & 0.92 & \textbf{0.95} & 0.92\\
\arrayrulecolor{black!60}\cmidrule(rr){2-12}
& Mean\tnote{$b$} & 0.71 & 0.58 & 0.59 & \underline{0.84} && 0.62 & 0.66 & 0.67 & \textbf{0.94} & 0.82\\
\arrayrulecolor{black!60}\midrule
\multirow{10}{*}{\rotatebox[origin=c]{90}{Objets}} & Bottle & 0.72 & 0.41 & 0.71 & 0.82 && \textbf{0.98} & \underline{0.97} & 0.77 & \textbf{0.98} & 0.94\\
& Cable & 0.64 & 0.48 & 0.52 & \underline{0.87} && 0.70 & 0.77 & 0.55 & \textbf{0.89} & 0.83\\
& Capsule & 0.55 & 0.49 & 0.44 & \underline{0.71} && \textbf{0.74} & 0.64 & 0.60 & \textbf{0.74} & 0.68\\
& Hazelnut & 0.83 & 0.60 & 0.68 & \underline{0.90} && \underline{0.90} & 0.88 & 0.85 & \textbf{0.94} & 0.86\\
& Metal Nut & 0.38 & 0.33 & 0.41 & 0.62 && 0.57 & 0.59 & 0.24 & \textbf{0.73} & \underline{0.67}\\
& Pill & 0.63 & 0.48 & 0.55 & 0.62 && 0.76 & 0.76 & 0.70 & \textbf{0.84} & \underline{0.79}\\
& Screw & 0.45 & 0.77 & 0.13 & \underline{0.80} && 0.68 & 0.60 & 0.30 & 0.74 & \textbf{1.00}\\
& Toothbrush & 0.36 & 0.44 & 0.51 & \underline{0.99} && 0.93 & 0.96 & 0.78 & \textbf{1.00} &\textbf{1.00}\\
& Transistor & 0.67 & 0.59 & 0.55 & \underline{0.90} && 0.84 & 0.85 & 0.46 & \textbf{0.91} &0.84\\
& Zipper & 0.44 & 0.41 & 0.70 & \underline{0.93} && 0.90 & 0.88 & 0.72 & \textbf{0.94} & 0.80\\
\arrayrulecolor{black!60}\cmidrule(rr){2-12}
& Mean\tnote{$c$} & 0.57 & 0.50 & 0.52 & 0.82 && 0.80 & 0.79 & 0.60 & \textbf{0.87} & \underline{0.84}\\
\arrayrulecolor{black!60}\midrule
& Global mean\tnote{$d$}  & 0.62 & 0.53 & 0.54 & 0.83 && 0.74 & 0.75 & 0.62 & \textbf{0.89} & \underline{0.84}\\
\arrayrulecolor{black}\bottomrule

\end{tabularx}

\begin{tablenotes}
\item [$a$] For each row, the best performing approach is highlighted in boldface and the second best is underlined.
\item [$b$] Mean AUC obtained over the classes of images belonging to the texture categories.
\item [$c$] Mean AUC obtained over the classes of images belonging to the object categories.
\item [$d$] Mean AUC obtained over the entire dataset.
\end{tablenotes}
\vspace*{0.4cm}
\end{threeparttable}
\end{table}

%% file: Tables/RESULT_pixel-wise.tex
\tabcolsep=0.075cm
\begin{table}[!h]
\begin{threeparttable}[b]
\caption{Pixel-wise detection AUC obtained with the residual- and uncertainty-based detection methods\tnote{\lowercase{$a$}}. }
\label{table:summary_pixel-wise}
\scriptsize

\newcommand\len{0.55}

\begin{tabularx}{8.9cm}{@{}llp{\len cm}p{\len cm}p{\len cm}p{\len cm}p{0.15cm}p{\len cm}p{\len cm}p{\len cm}p{\len cm}p{1.2cm}@{}} \toprule

&& \multicolumn{4}{c}{\textbf{Uncertainty}} && \multicolumn{5}{c}{\textbf{Residual}}  \\
\arrayrulecolor{black!60}\cmidrule(rr){3-6}\cmidrule(rr){8-12}
\multicolumn{2}{l}{Network} & \multicolumn{2}{c}{\textbf{AE}} & \multicolumn{2}{c}{\textbf{AESc}} && \multicolumn{2}{c}{\textbf{AE}} & \multicolumn{2}{c}{\textbf{AESc}} & \textbf{AE\textsubscript{L2} \cite{Bergmann2019}} \\
\arrayrulecolor{black!60}\cmidrule(rr){3-4}\cmidrule(rr){5-6}\cmidrule(rr){8-9}\cmidrule(rr){10-11}
\multicolumn{2}{l}{Corruption} & None & Stain & None & Stain && None & Stain & None & Stain &  \\

\arrayrulecolor{black!60}\midrule\midrule

 \multirow{5}{*}{\rotatebox[origin=c]{90}{Textures}} & Carpet & 0.55 & 0.54 & 0.43 & \textbf{0.91} && 0.57 & 0.62 & 0.52 & \underline{0.79} & 0.59\\
& Grid & 0.52 & 0.49 & 0.50 & \textbf{0.95} && 0.81 & 0.82 & 0.57 & 0.89 & \underline{0.90}\\
& Leather &  0.86 & 0.52 & 0.58 & \underline{0.87} && 0.79 & 0.82 & 0.71 & \textbf{0.95} & 0.75 \\ 
& Tile & 0.54 & 0.50 & 0.53 & \textbf{0.79} && 0.45 & 0.54 & 0.62 & \underline{0.74} & 0.51\\
& Wood &0.61 & 0.48 & 0.51 & \textbf{0.84} && 0.64 & 0.71 & 0.65 & \textbf{0.84} & \underline{0.73}\\
\arrayrulecolor{black!60}\cmidrule(rr){2-12}
& Mean\tnote{$b$} & 0.62 & 0.51 & 0.51 & \textbf{0.87} && 0.65 & 0.70 & 0.61 & \underline{0.84}  & 0.70\\
\arrayrulecolor{black!60}\midrule

 \multirow{10}{*}{\rotatebox[origin=c]{90}{Objects}} & Bottle & 0.68 & 0.63 & 0.64 & \textbf{0.88} && 0.85 & \textbf{0.88} & 0.47 & 0.84 & \underline{0.86} \\
& Cable & 0.54 & 0.70 & 0.66 & 0.84 && 0.62 & 0.83 & 0.72 & \underline{0.85} & \textbf{0.86}\\
& Capsule & \underline{0.92} & 0.89 & 0.65 & \textbf{0.93} && 0.87 & 0.87 & 0.63 & 0.83  & 0.88 \\
& Hazelnut & \textbf{0.95} & 0.91 & 0.60 & 0.89 && 0.92 & \underline{0.93} & 0.79 & 0.88 & \textbf{0.95} \\
& Metal Nut & 0.79 & 0.73 & 0.50 & 0.62 &&  0.82 & \underline{0.84} & 0.52 & 0.57 & \textbf{0.86} \\
& Pill & \underline{0.82} & \underline{0.82} & 0.61 & \textbf{0.85} && 0.81 & 0.81 & 0.64 & 0.74 & \textbf{0.85}\\
& Screw & 0.94 & 0.94 & 0.61 & \underline{0.95} && 0.93 & 0.93 & 0.72 & 0.86 & \textbf{0.96} \\
& Toothbrush & 0.84 & 0.83 & 0.79 & \textbf{0.93} && \underline{0.92} & \textbf{0.93} & 0.73 & \textbf{0.93} & \textbf{0.93}\\
& Transistor & 0.79 & 0.64 & 0.51 & 0.78 && 0.79 & \underline{0.82} & 0.56 & 0.80 & \textbf{0.86}\\
& Zipper & \underline{0.78} & 0.77 & 0.60 & \textbf{0.90} && 0.73 & 0.75 & 0.60 & \underline{0.78} & 0.77\\
\arrayrulecolor{black!60}\cmidrule(rr){2-12}
& Mean\tnote{$c$} &  0.81 & 0.79 & 0.62 & \underline{0.86} && 0.83 & \underline{0.86} & 0.64 & 0.81 & \textbf{0.88} \\
\arrayrulecolor{black!60}\midrule

& Global mean\tnote{$d$} & 0.74 & 0.69 & 0.58 & \textbf{0.86} && 0.77 & 0.81 & 0.63 & \underline{0.82}  & \underline{0.82} \\
\arrayrulecolor{black}\bottomrule

\end{tabularx}
\begin{tablenotes}
\item [$a$] For each row, the best performing approach is highlighted in boldface and the second best is underlined.
\item [$b$] Mean AUC obtained over the classes of images belonging to the texture categories.
\item [$c$] Mean AUC obtained over the classes of images belonging to the object categories.
\item [$d$] Mean AUC obtained over the entire dataset.
\end{tablenotes}
\end{threeparttable}

\vspace*{-0.2cm}
\end{table}

%% file: imgs/Results/residualVSuncertainty.tex
{
\newcommand\inputONEPATH{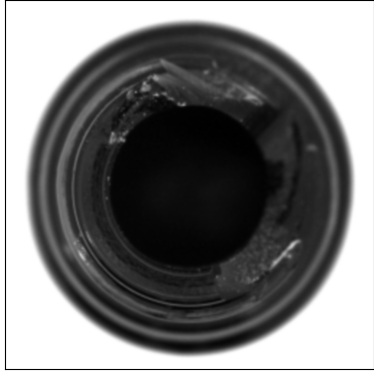}
\newcommand\maskONEPATH{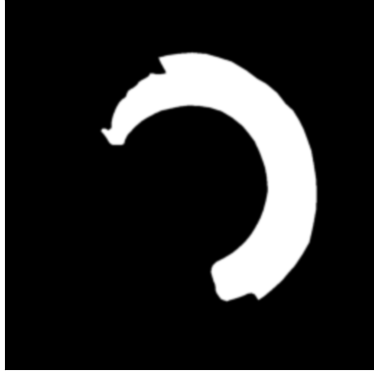}
\newcommand\predONEPATH{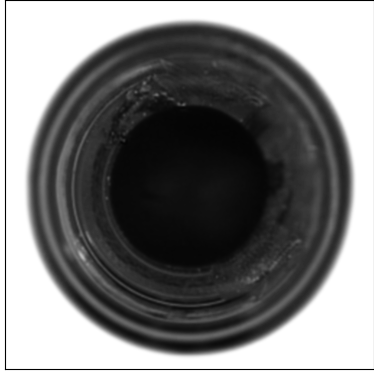}
\newcommand\residualONEPATH{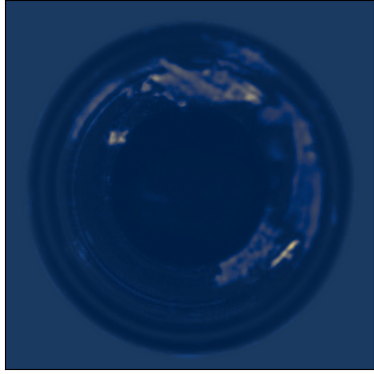}
\newcommand\uncertaintyONEPATH{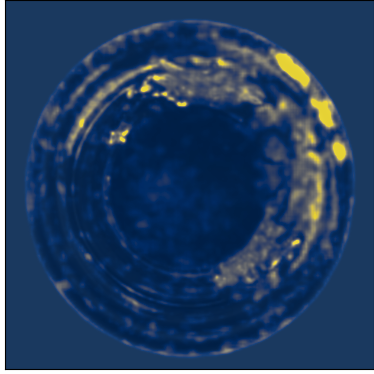}
\newcommand\colorbarUncertaintyONEPATH{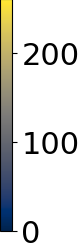}

\newcommand\inputTHREEPATH{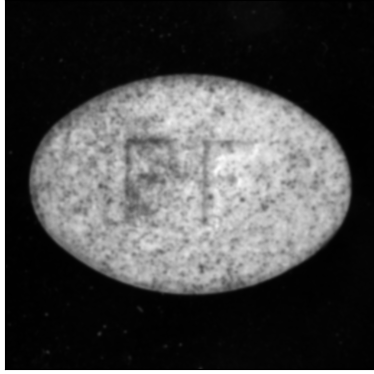}
\newcommand\maskTHREEPATH{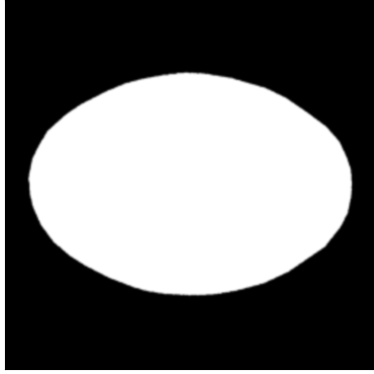}
\newcommand\predTHREEPATH{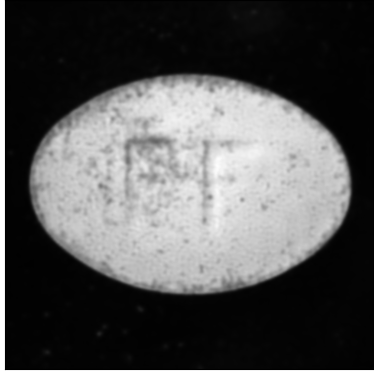}
\newcommand\residualTHREEPATH{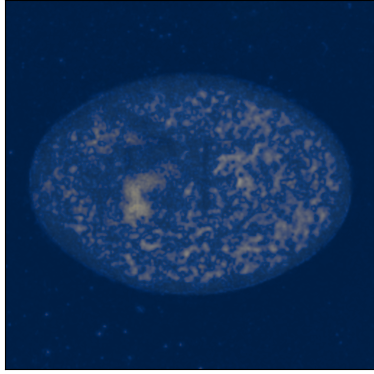}
\newcommand\uncertaintyTHREEPATH{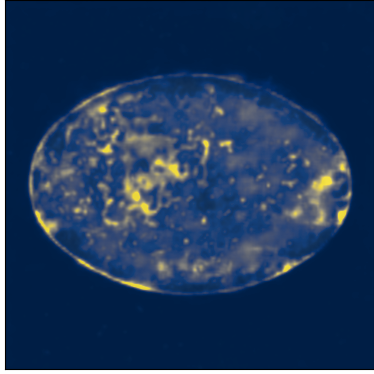}
\newcommand\colorbarUncertaintyTHREEPATH{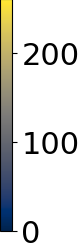}

\newcommand\inputTWOPATH{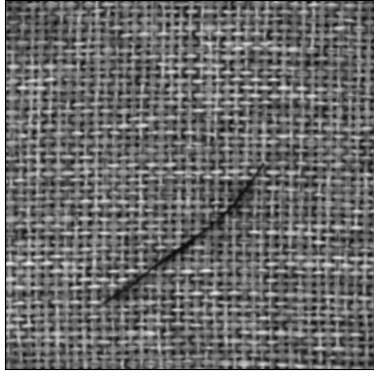}
\newcommand\maskTWOPATH{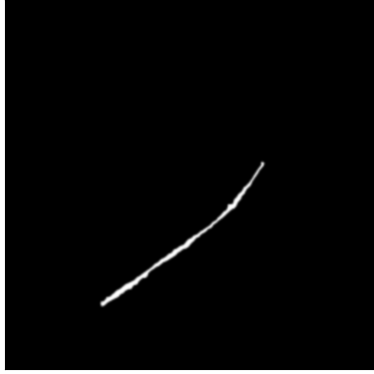}
\newcommand\predTWOPATH{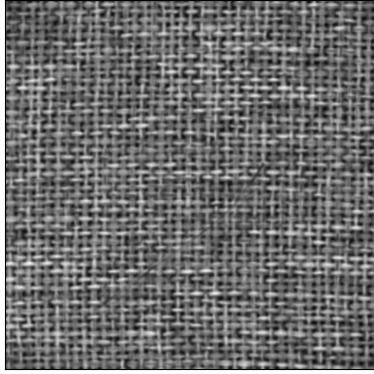}
\newcommand\residualTWOPATH{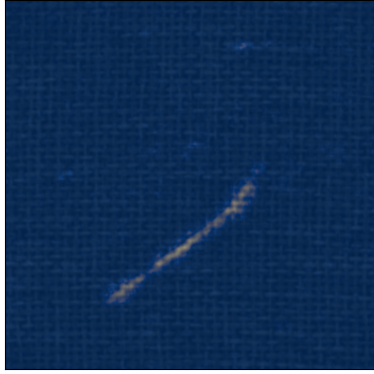}
\newcommand\uncertaintyTWOPATH{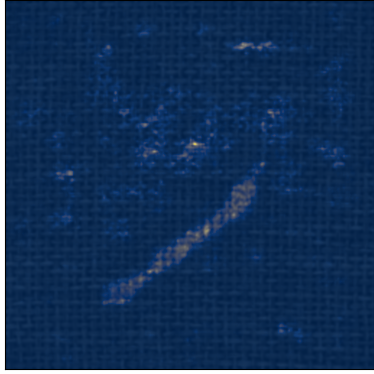}
\newcommand\colorbarUncertaintyTWOPATH{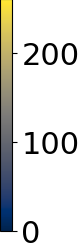}

\newcommand\inputFOURPATH{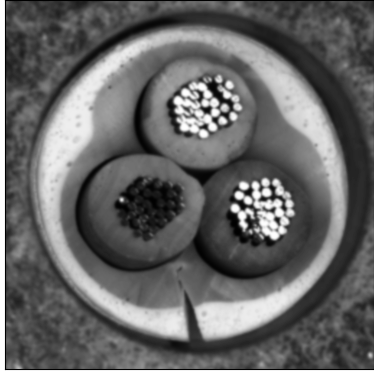}
\newcommand\maskFOURPATH{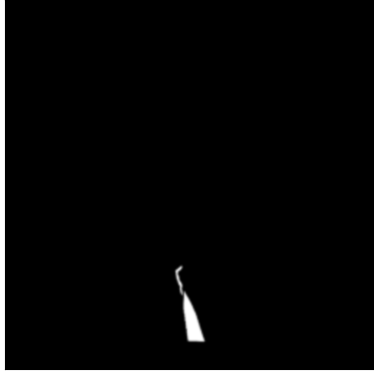}
\newcommand\predFOURPATH{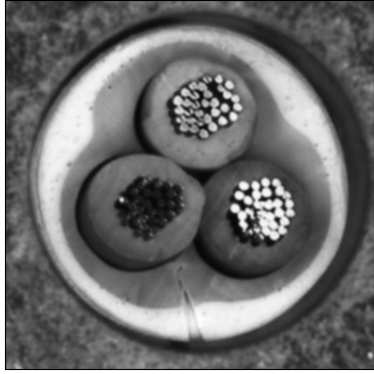}
\newcommand\residualFOURPATH{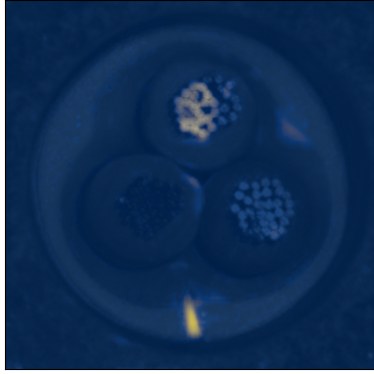}
\newcommand\uncertaintyFOURPATH{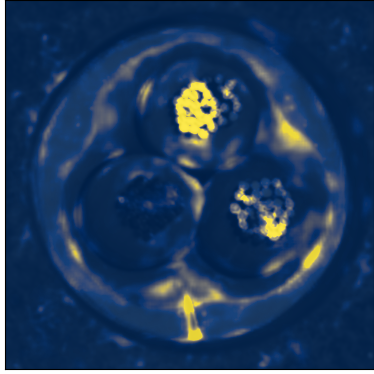}
\newcommand\colorbarUncertaintyFOURPATH{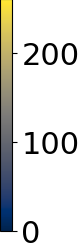}

\begin{tikzpicture}[node distance=0.5cm,>=latex', every text node part/.style={align=center}]

	\newcommand\xidt{0.2}
	\newcommand\imsize{1.5}
	\newcommand\vertshift{0.85}
	\newcommand\mysize{0.4}
	\newcommand\yidt{-3.4}
	\newcommand\xindent{9}

	\node at (0,\vertshift+1.0) [fontscale=\mysize] {Mask};
	\node at (\imsize+1*\xidt,\vertshift+1.0) [fontscale=\mysize] {Input};
	\node at (2*\imsize+2*\xidt,\vertshift+1.0) [fontscale=\mysize] {Prediction};
	\node at (3*\imsize+3*\xidt,\vertshift+1.0) [fontscale=\mysize] {Residual};
	\node at (4*\imsize+4*\xidt,\vertshift+1.0) [fontscale=\mysize] {Uncertainty};
	
	

	\node at (0,\vertshift) [fontscale=\mysize, text width=1.5cm] {\includegraphics[width=\imsize cm]{\maskONEPATH}};
	\node at (\imsize+1*\xidt,\vertshift) [fontscale=\mysize, text width=1.5cm] {\includegraphics[width=\imsize cm]{\inputONEPATH}};
	\node at (2*\imsize+2*\xidt,\vertshift) [fontscale=\mysize, text width=1.5cm] {\includegraphics[width=\imsize cm]{\predONEPATH}};
	\node at (3*\imsize+3*\xidt,\vertshift) [fontscale=\mysize, text width=1.5cm] {\includegraphics[width=\imsize cm]{\residualONEPATH}};
	\node at (4*\imsize+4*\xidt,\vertshift) [fontscale=\mysize, text width=1.5cm] {\includegraphics[width=\imsize cm]{\uncertaintyONEPATH}};
	\node at (5*\imsize+2*\xidt,\vertshift-0.05) [fontscale=\mysize, text width=1.5cm] {\includegraphics[height=1.5cm]{\colorbarUncertaintyONEPATH}};
	
	\node at (0,-\vertshift) [fontscale=\mysize, text width=1.5cm] {\includegraphics[width=\imsize cm]{\maskTHREEPATH}};
	\node at (\imsize+1*\xidt,-\vertshift) [fontscale=\mysize, text width=1.5cm] {\includegraphics[width=\imsize cm]{\inputTHREEPATH}};
	\node at (2*\imsize+2*\xidt,-\vertshift) [fontscale=\mysize, text width=1.5cm] {\includegraphics[width=\imsize cm]{\predTHREEPATH}};
	\node at (3*\imsize+3*\xidt,-\vertshift) [fontscale=\mysize, text width=1.5cm] {\includegraphics[width=\imsize cm]{\residualTHREEPATH}};
	\node at (4*\imsize+4*\xidt,-\vertshift) [fontscale=\mysize, text width=1.5cm] {\includegraphics[width=\imsize cm]{\uncertaintyTHREEPATH}};
	\node at (5*\imsize+2*\xidt,-\vertshift-0.05) [fontscale=\mysize, text width=1.5cm] {\includegraphics[height=1.5cm]{\colorbarUncertaintyTHREEPATH}};

	\node at (0,-3*\vertshift) [fontscale=\mysize, text width=1.5cm] {\includegraphics[width=\imsize cm]{\maskTWOPATH}};
	\node at (\imsize+1*\xidt,-3*\vertshift) [fontscale=\mysize, text width=1.5cm] {\includegraphics[width=\imsize cm]{\inputTWOPATH}};
	\node at (2*\imsize+2*\xidt,-3*\vertshift) [fontscale=\mysize, text width=1.5cm] {\includegraphics[width=\imsize cm]{\predTWOPATH}};
	\node at (3*\imsize+3*\xidt,-3*\vertshift) [fontscale=\mysize, text width=1.5cm] {\includegraphics[width=\imsize cm]{\residualTWOPATH}};
	\node at (4*\imsize+4*\xidt,-3*\vertshift) [fontscale=\mysize, text width=1.5cm] {\includegraphics[width=\imsize cm]{\uncertaintyTWOPATH}};
	\node at (5*\imsize+2*\xidt,-3*\vertshift-0.05) [fontscale=\mysize, text width=1.5cm] {\includegraphics[height=1.5cm]{\colorbarUncertaintyTWOPATH}};
	
	\node at (0,-5*\vertshift) [fontscale=\mysize, text width=1.5cm] {\includegraphics[width=\imsize cm]{\maskFOURPATH}};
	\node at (\imsize+1*\xidt,-5*\vertshift) [fontscale=\mysize, text width=1.5cm] {\includegraphics[width=\imsize cm]{\inputFOURPATH}};
	\node at (2*\imsize+2*\xidt,-5*\vertshift) [fontscale=\mysize, text width=1.5cm] {\includegraphics[width=\imsize cm]{\predFOURPATH}};
	\node at (3*\imsize+3*\xidt,-5*\vertshift) [fontscale=\mysize, text width=1.5cm] {\includegraphics[width=\imsize cm]{\residualFOURPATH}};
	\node at (4*\imsize+4*\xidt,-5*\vertshift) [fontscale=\mysize, text width=1.5cm] {\includegraphics[width=\imsize cm]{\uncertaintyFOURPATH}};
	\node at (5*\imsize+2*\xidt,-5*\vertshift-0.05) [fontscale=\mysize, text width=1.5cm] {\includegraphics[height=1.5cm]{\colorbarUncertaintyFOURPATH}};

	\end{tikzpicture}
	
}

%% file: imgs/Results/corruptionComparison_BottleReal60.tex
{ 

\newif\ifMC
    \MCfalse  
    
\newif\iflegend
    \legendtrue  

\newcommand\inputPATH{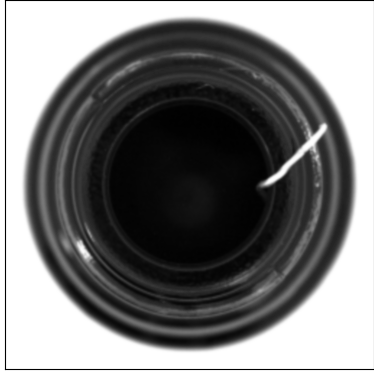}
\newcommand\maskPATH{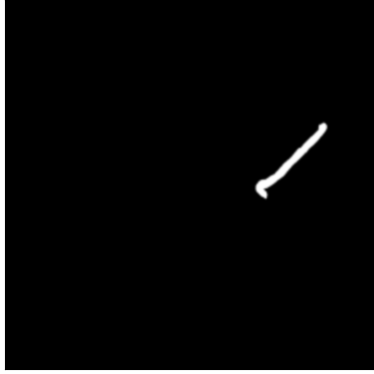}
\newcommand\colorbarResidualPATH{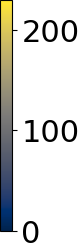}
\newcommand\colorbarUncertaintyPATH{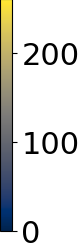}

\newcommand\CAESSCpredPATH{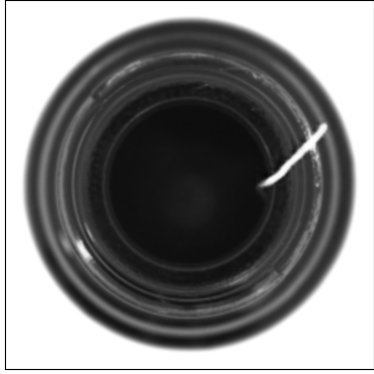}
\newcommand\CAESSCresidualPATH{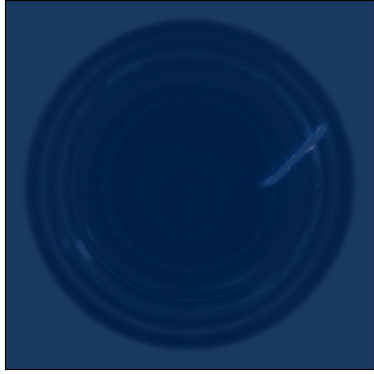}
\newcommand\CAESSCuncertaintyPATH{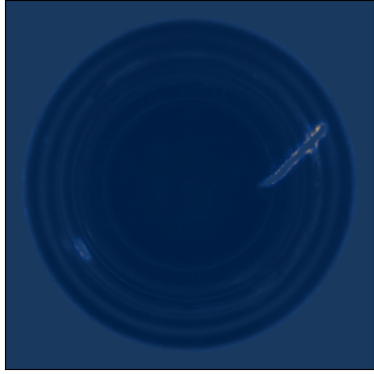}

\newcommand\stainCAESSCpredPATH{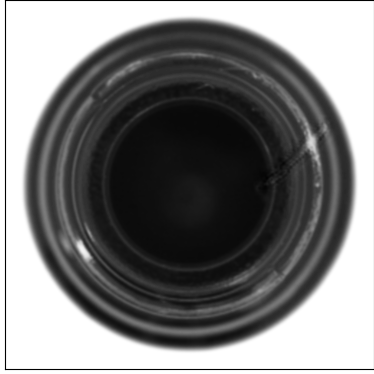}
\newcommand\stainCAESSCresidualPATH{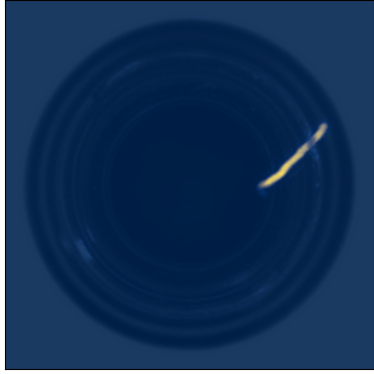}
\newcommand\stainCAESSCuncertaintyPATH{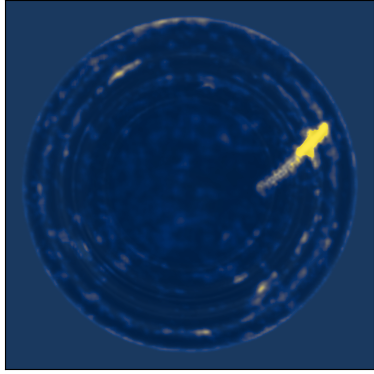}

\newcommand\dropsCAESSCpredPATH{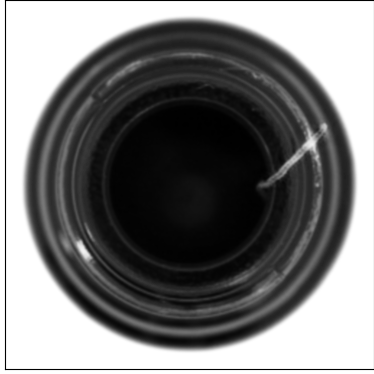}
\newcommand\dropsCAESSCresidualPATH{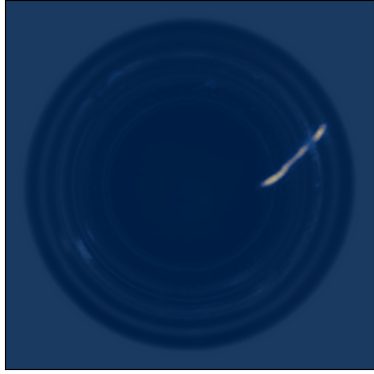}
\newcommand\dropsCAESSCuncertaintyPATH{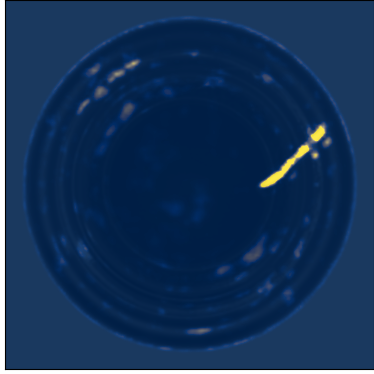}

\newcommand\scratchCAESSCpredPATH{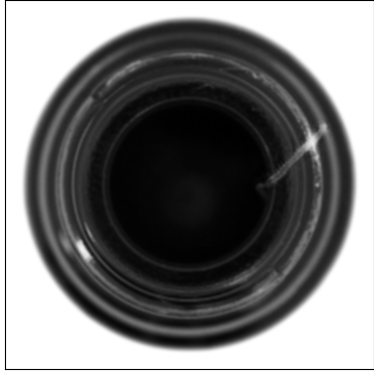}
\newcommand\scratchCAESSCresidualPATH{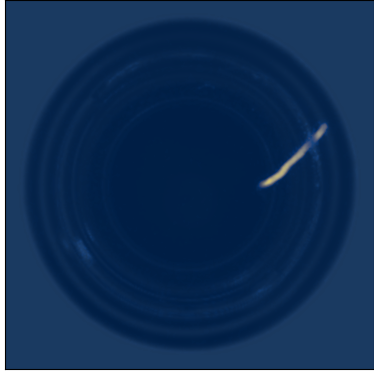}
\newcommand\scratchCAESSCuncertaintyPATH{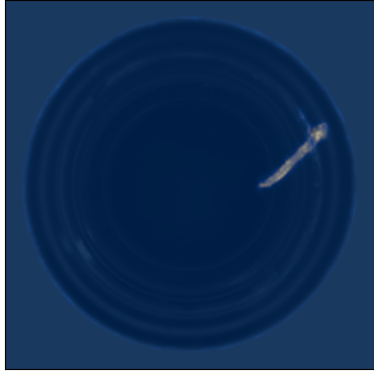}

\newcommand\structuralCAESSCpredPATH{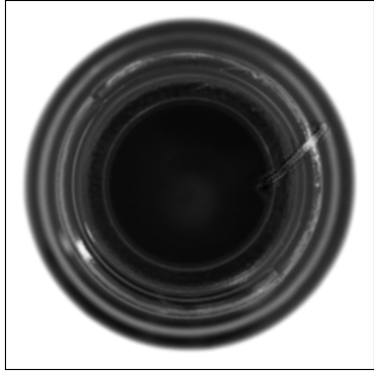}
\newcommand\structuralCAESSCresidualPATH{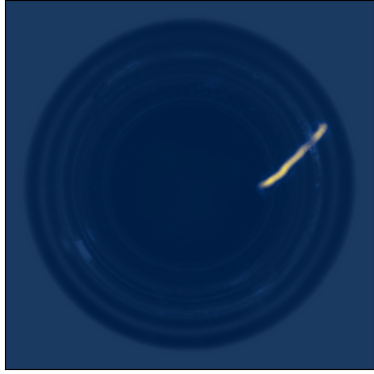}
\newcommand\structuralCAESSCuncertaintyPATH{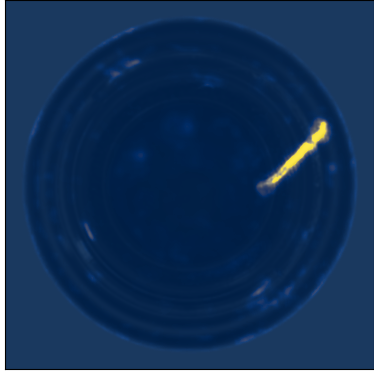}

\newcommand\stainGaussianCAESSCpredPATH{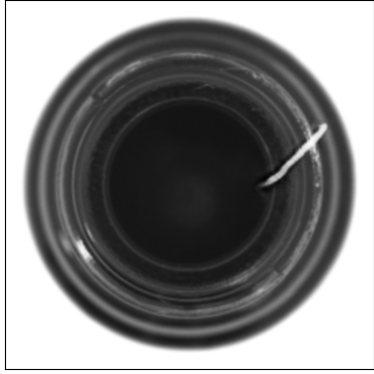}
\newcommand\stainGaussianCAESSCresidualPATH{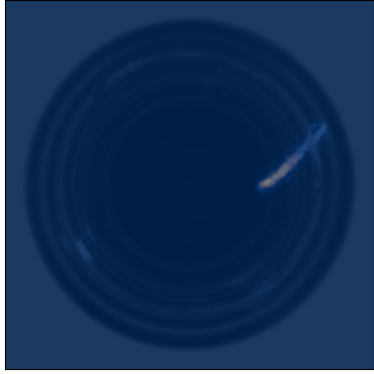}
\newcommand\stainGaussianCAESSCuncertaintyPATH{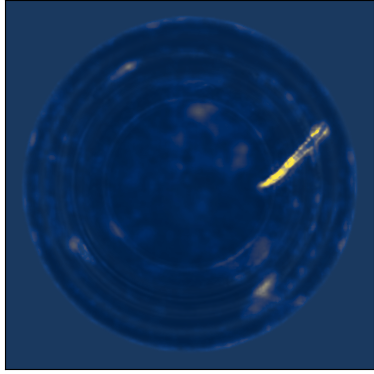}

\newcommand\gaussianTENCAESSCpredPATH{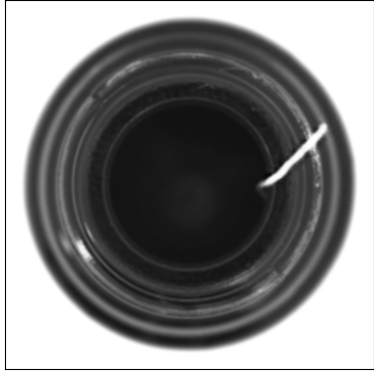}
\newcommand\gaussianTENCAESSCresidualPATH{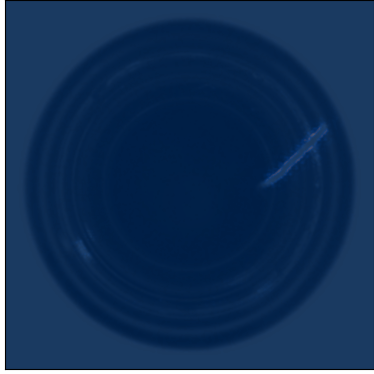}
\newcommand\gaussianTENCAESSCuncertaintyPATH{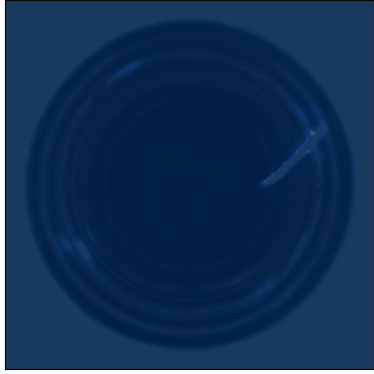}

\newcommand\gaussianTWENTYCAESSCpredPATH{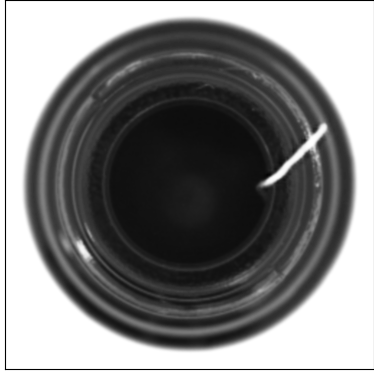}
\newcommand\gaussianTWENTYCAESSCresidualPATH{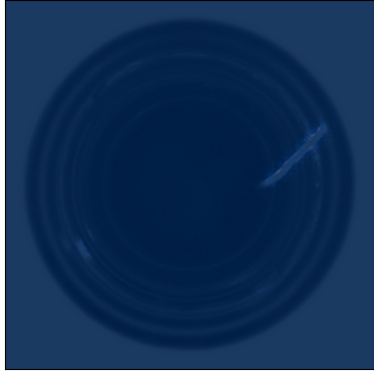}
\newcommand\gaussianTWENTYCAESSCuncertaintyPATH{imgs/Prediction_analysis/Bottle_Gaussian20256x256_ImtoIm_CAeSSC2_d6_f5_n16_LR0.01RedOnPlateau_B16_Ep250_Gray_NoDA/MC_Real_60_VARIANCEMC.png}

\newcommand\gaussianFORTYCAESSCpredPATH{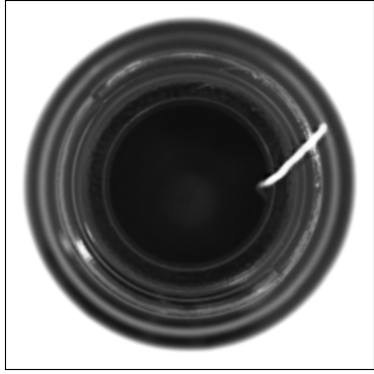}
\newcommand\gaussianFORTYCAESSCresidualPATH{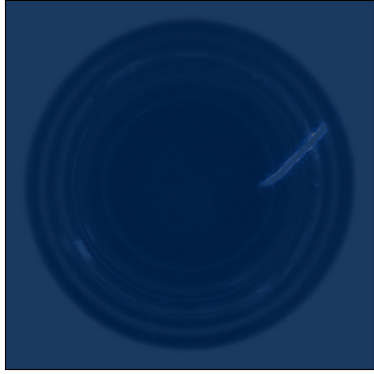}
\newcommand\gaussianFORTYCAESSCuncertaintyPATH{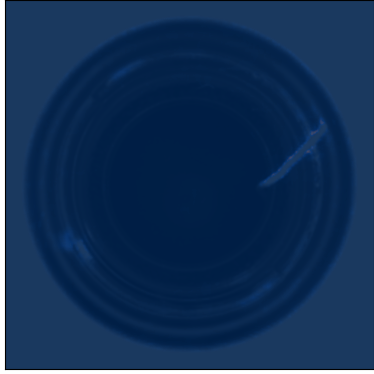}

\newcommand\gaussianEIGHTYCAESSCpredPATH{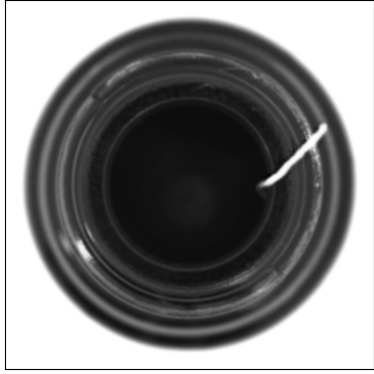}
\newcommand\gaussianEIGHTYCAESSCresidualPATH{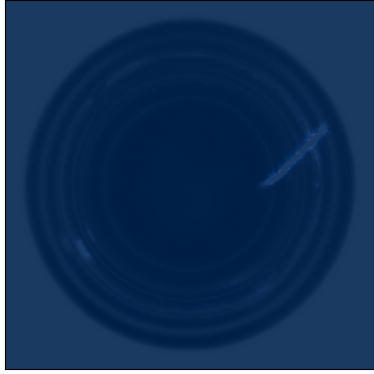}
\newcommand\gaussianEIGHTYCAESSCuncertaintyPATH{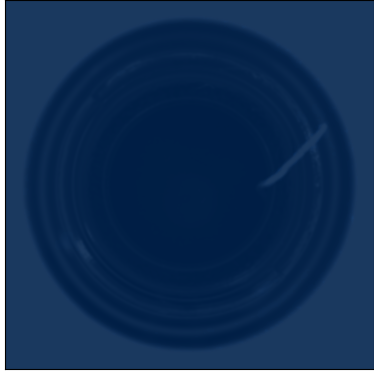}

\includestandalone[width=12cm]{imgs/Results/corruptionComparison}

}

%% file: imgs/Results/corruptionComparison_WoodReal8.tex
{ 

\newif\ifMC
    \MCfalse  
    
\newif\iflegend
    \legendfalse  

\newcommand\inputPATH{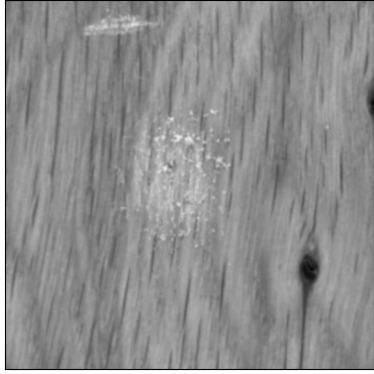}
\newcommand\maskPATH{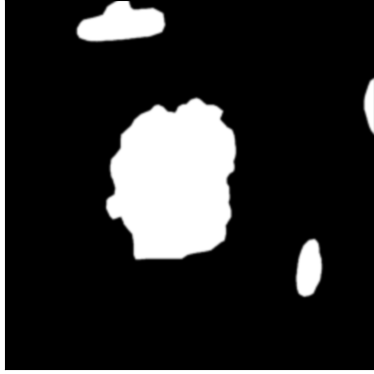}
\newcommand\colorbarResidualPATH{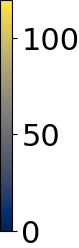}
\newcommand\colorbarUncertaintyPATH{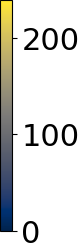}

\newcommand\CAESSCpredPATH{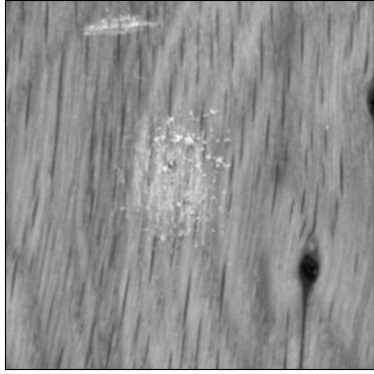}
\newcommand\CAESSCresidualPATH{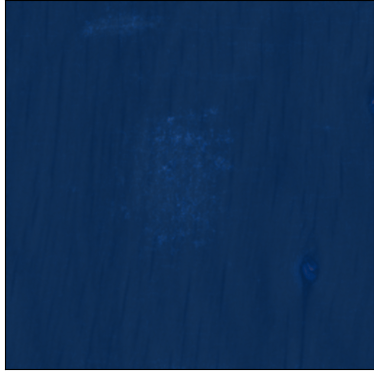}
\newcommand\CAESSCuncertaintyPATH{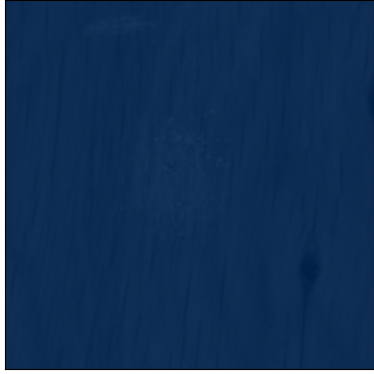}

\newcommand\stainCAESSCpredPATH{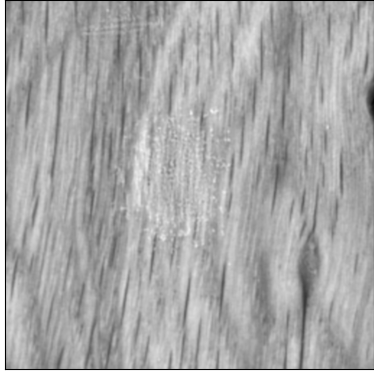}
\newcommand\stainCAESSCresidualPATH{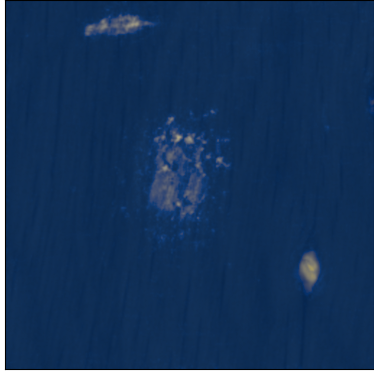}
\newcommand\stainCAESSCuncertaintyPATH{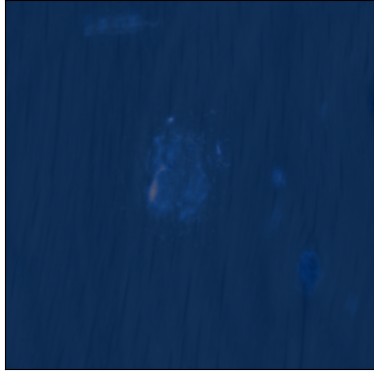}

\newcommand\dropsCAESSCpredPATH{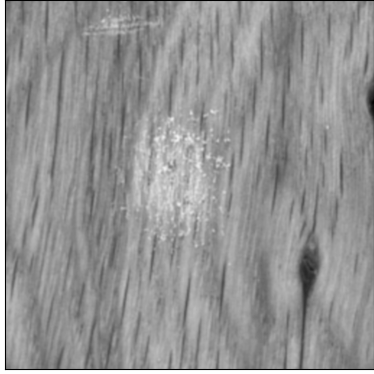}
\newcommand\dropsCAESSCresidualPATH{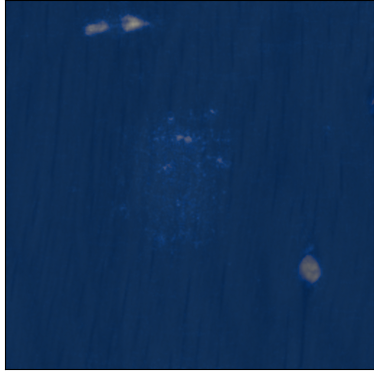}
\newcommand\dropsCAESSCuncertaintyPATH{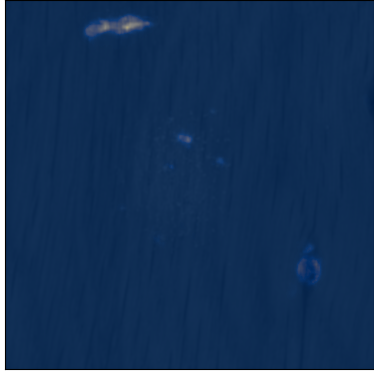}

\newcommand\scratchCAESSCpredPATH{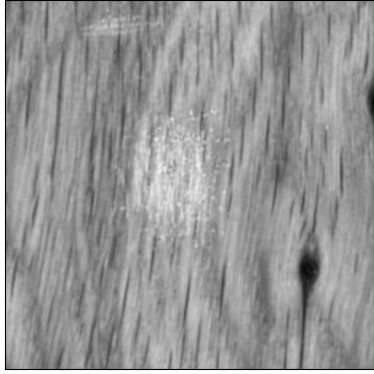}
\newcommand\scratchCAESSCresidualPATH{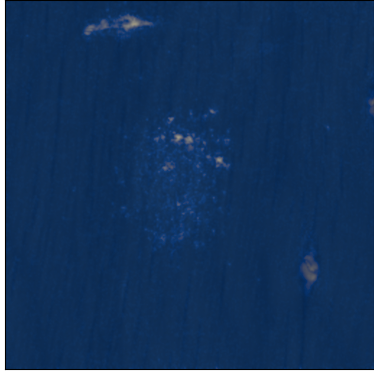}
\newcommand\scratchCAESSCuncertaintyPATH{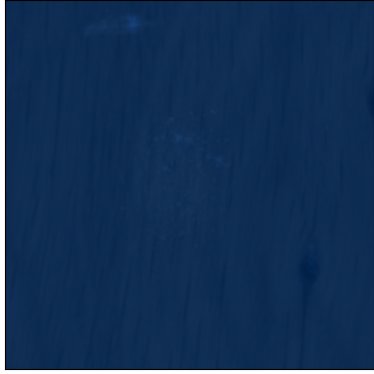}

\newcommand\structuralCAESSCpredPATH{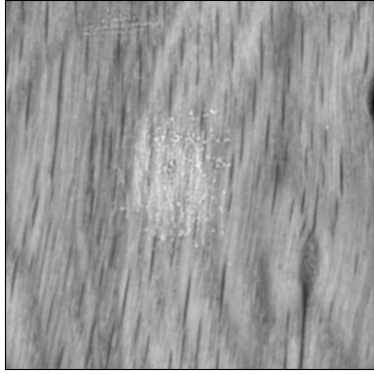}
\newcommand\structuralCAESSCresidualPATH{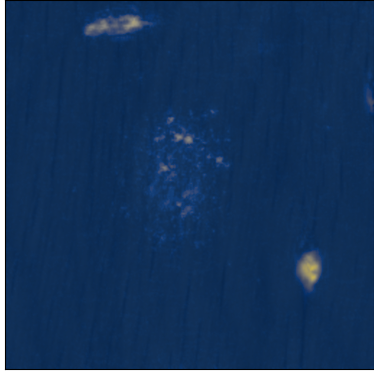}
\newcommand\structuralCAESSCuncertaintyPATH{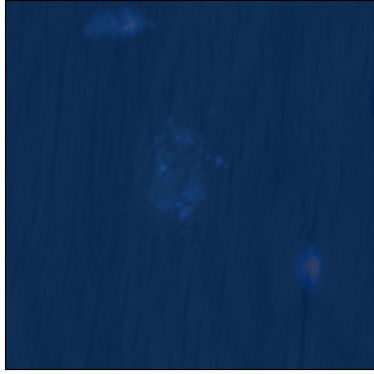}

\newcommand\stainGaussianCAESSCpredPATH{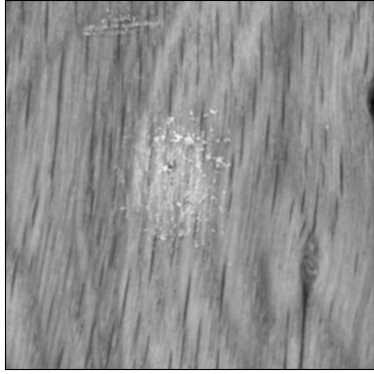}
\newcommand\stainGaussianCAESSCresidualPATH{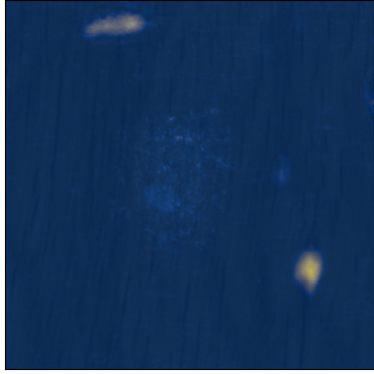}
\newcommand\stainGaussianCAESSCuncertaintyPATH{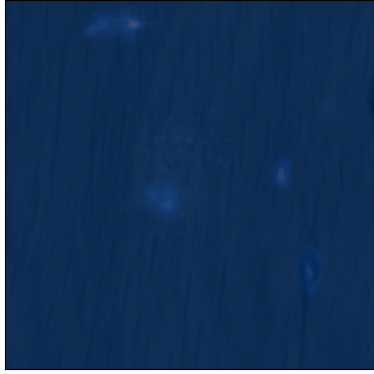}

\newcommand\gaussianTENCAESSCpredPATH{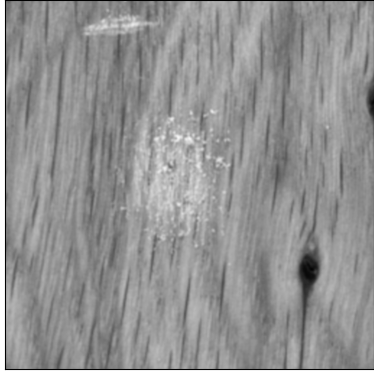}
\newcommand\gaussianTENCAESSCresidualPATH{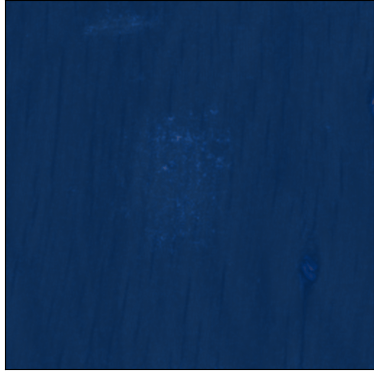}
\newcommand\gaussianTENCAESSCuncertaintyPATH{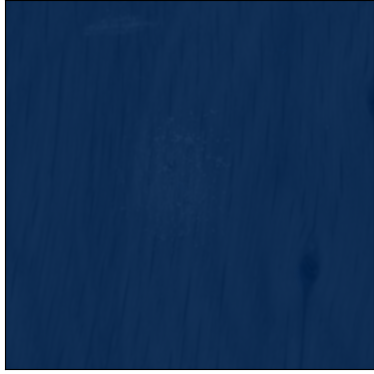}

\newcommand\gaussianTWENTYCAESSCpredPATH{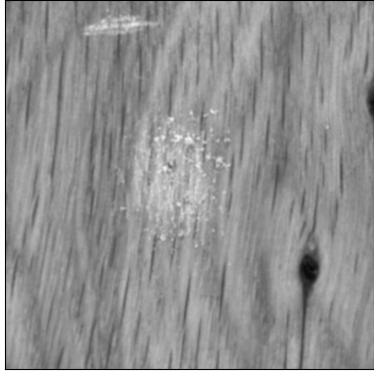}
\newcommand\gaussianTWENTYCAESSCresidualPATH{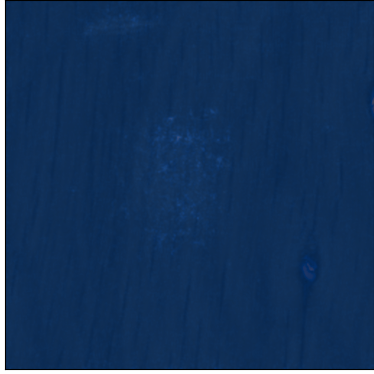}
\newcommand\gaussianTWENTYCAESSCuncertaintyPATH{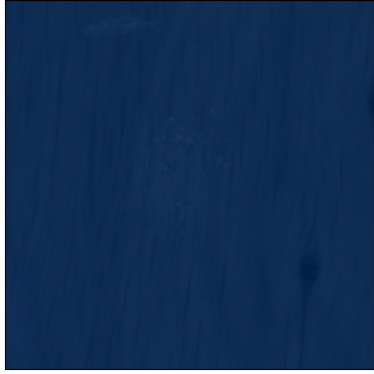}

\newcommand\gaussianFORTYCAESSCpredPATH{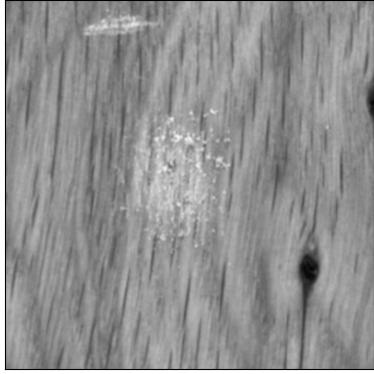}
\newcommand\gaussianFORTYCAESSCresidualPATH{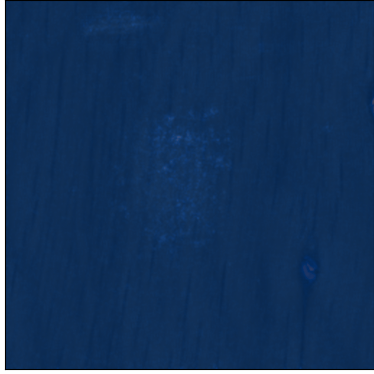}
\newcommand\gaussianFORTYCAESSCuncertaintyPATH{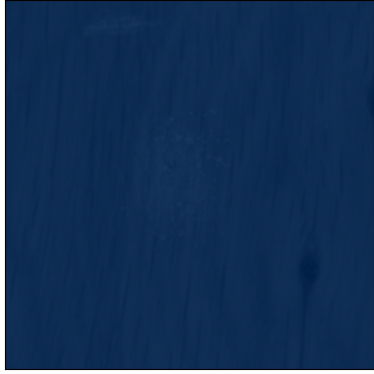}

\newcommand\gaussianEIGHTYCAESSCpredPATH{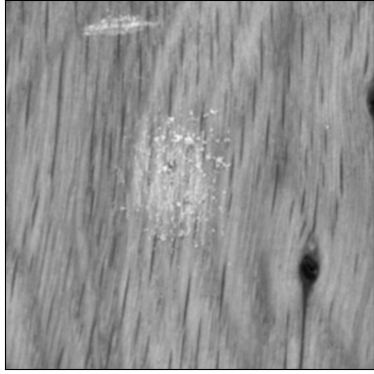}
\newcommand\gaussianEIGHTYCAESSCresidualPATH{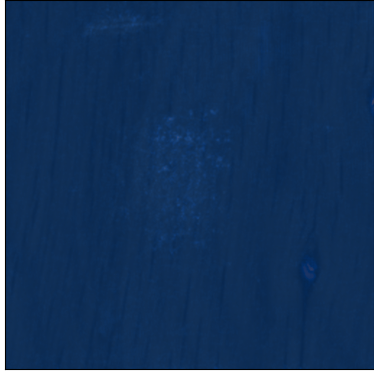}
\newcommand\gaussianEIGHTYCAESSCuncertaintyPATH{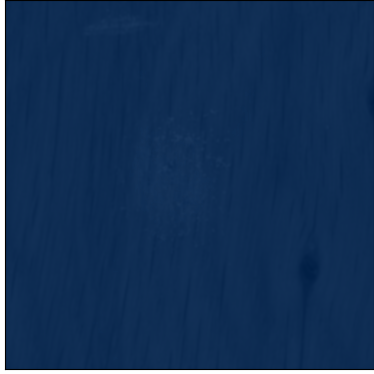}

\includestandalone[width=12cm]{imgs/Results/corruptionComparison}

}

%% file: Tables/APPENDIX_image-wise.tex
\tabcolsep=0.03cm

\begin{table}[!t]
\begin{threeparttable}[b]
\caption{Image-wise AUC obtained with the AESc network trained with different noise models with the residual-based problem formulation\tnote{$a$}. }
\label{table:AESc-residual-image}
\centering
\scriptsize

\newcommand\len{0.75}
\newcommand\lenbis{0.55}

\begin{tabularx}{9.0cm}{@{}ll p{\len cm}p{\len cm}p{\lenbis cm}p{\lenbis cm}p{\lenbis cm}p{\len cm}p{\len cm}p{\len cm}p{\len cm}p{0.7 cm}@{}} \toprule

&\multirow{2}{*}{\hspace*{-0.2cm}Corruption} & \multirow{2}{*}{None} & \multirow{2}{*}{Drops} & \multicolumn{4}{c}{Gaussian noise ($\sigma$)} & \multirow{2}{*}{Scratch} & \multirow{2}{*}{Stain} & \multirow{2}{*}{Mix1} & \multirow{2}{*}{Mix2} \\

\arrayrulecolor{black!60}\cmidrule(rr){5-8}  

 & &\multicolumn{2}{c}{} & \textbf{$0.1$} & \textbf{$0.2$} & \textbf{$0.4$} & \textbf{$0.8$} & & & &  \\ 

\arrayrulecolor{black!60}\midrule\midrule

\multirow{5}{*}{\rotatebox[origin=c]{90}{Textures}} & Carpet & 0.48 & \underline{0.87} & 0.52 & 0.46 & 0.51 & 0.53 & 0.63 & \textbf{0.89} & 0.84 & 0.84 \\
& Grid & 0.52 & 0.94 & 0.55 & 0.69 & 0.59 & 0.72 & 0.79 & \textbf{0.97} & 0.91 & \underline{0.96} \\
& Leather & 0.56 & 0.87 & 0.72 & 0.71 & 0.74 & 0.71 & 0.77 & \textbf{0.89} & \underline{0.88} & \textbf{0.89} \\
& Tile & 0.88 & 0.94 & 0.94 & 0.92 & 0.90 & 0.92 & 0.95 & \textbf{0.99} & \underline{0.98} & 0.96 \\
& Wood & 0.92 & \textbf{0.99} & 0.89 & 0.90 & 0.91 & 0.85 & \underline{0.96} & 0.95 & 0.94 & 0.79 \\
\arrayrulecolor{black!60}\cmidrule(rr){2-12}
& Mean\tnote{$b$} & 0.67 & \underline{0.92} & 0.72 & 0.74 & 0.73 & 0.75 & 0.82 & \textbf{0.94} & 0.91 & 0.89 \\
\arrayrulecolor{black!60}\midrule

\multirow{10}{*}{\rotatebox[origin=c]{90}{Objets}} & Bottle & 0.77 & \textbf{0.99} & 0.82 & 0.85 & 0.81 & 0.75 & 0.91 & \underline{0.98} & \underline{0.98} & 0.97 \\
& Cable & 0.55 & 0.60 & 0.58 & 0.53 & 0.49 & 0.46 & 0.60 & \underline{0.89} & 0.87 & \textbf{0.90} \\
& Capsule & 0.60 & \underline{0.71} & 0.58 & 0.68 & 0.57 & 0.59 & 0.66 & \textbf{0.74} & \textbf{0.74} & 0.53 \\
& Hazelnut & 0.85 & \textbf{0.98} & 0.75 & 0.73 & 0.92 & 0.73 & \underline{0.96} & 0.94 & 0.93 & 0.81 \\
& Metal Nut & 0.24 & 0.54 & 0.32 & 0.27 & 0.28 & 0.24 & 0.44 & \underline{0.73} & 0.71 & \textbf{0.86} \\
& Pill & 0.70 & \underline{0.79} & 0.69 & 0.71 & 0.73 & 0.68 & 0.78 & \textbf{0.84} & 0.77 & 0.78 \\
& Screw & 0.30 & 0.46 & \underline{0.91} & \textbf{0.99} & 0.78 & 0.65 & 0.71 & 0.74 & 0.22 & 0.72 \\
& Toothbrush & 0.78 & \textbf{1.00} & \underline{0.99} & 0.98 & 0.79 & 0.82 & 0.87 & \textbf{1.00} & \textbf{1.00} & \textbf{1.00} \\
& Transistor & 0.46 & 0.83 & 0.55 & 0.49 & 0.48 & 0.50 & 0.68 & \underline{0.91} & \textbf{0.92} & \textbf{0.92} \\
& Zipper & 0.72 & 0.93 & 0.66 & 0.63 & 0.69 & 0.58 & 0.79 & \underline{0.94} & 0.90 & \textbf{0.98} \\
\arrayrulecolor{black!60}\cmidrule(rr){2-12}
& Mean\tnote{$c$} & 0.60 & 0.78 & 0.68 & 0.69 & 0.65 & 0.60 & 0.74 & \textbf{0.87} & 0.80 & \underline{0.85} \\
\arrayrulecolor{black!60}\midrule
& Global mean\tnote{$d$} & 0.62 & 0.83 & 0.70 & 0.70 & 0.68 & 0.65 & 0.77 & \textbf{0.89} & 0.84 & \underline{0.86} \\
\arrayrulecolor{black}\bottomrule

\end{tabularx}

\begin{tablenotes}
\item [$a$] For each row, the best performing approach is highlighted in boldface and the second best is underlined.
\item [$b$] Mean AUC obtained over the classes of images belonging to the texture categories.
\item [$c$] Mean AUC obtained over the classes of images belonging to the object categories.
\item [$d$] Mean AUC obtained over the entire dataset.
\end{tablenotes}
\end{threeparttable}
\end{table}

%% file: Chapters/Conclusion.tex
In this work, we considered an anomaly detection method based on the reconstruction of a clean image from any arbitrary image. It builds on convolutional autoencoder and relies on the reconstruction residual or the prediction uncertainty, estimated with the Monte Carlo dropout technique, to detect anomalies. We demonstrated the benefits of considering an autoencoder architecture equipped with skip connections, as long as the training images are corrupted with our Stain noise model to avoid convergence towards an identity operator. This new approach performs significantly better than traditional autoencoders to detect real defects on texture images of the MVTec AD dataset.\\
Furthermore, we also provided a fair comparison between the residual- and uncertainty-based detection strategies relying on our AESc + Stain model. Unlike the reconstruction residual, the uncertainty indicator is independent of the contrast between the defect and its surroundings, which is particularly relevant for low contrast defects localization. However, in comparison to the residual-based detection strategy, the uncertainty-based approach increases the false positive rate in the clean structures.

%% file: main.bbl
\begin{thebibliography}{10}
\providecommand{\url}[1]{#1}
\csname url@samestyle\endcsname
\providecommand{\newblock}{\relax}
\providecommand{\bibinfo}[2]{#2}
\providecommand{\BIBentrySTDinterwordspacing}{\spaceskip=0pt\relax}
\providecommand{\BIBentryALTinterwordstretchfactor}{4}
\providecommand{\BIBentryALTinterwordspacing}{\spaceskip=\fontdimen2\font plus
\BIBentryALTinterwordstretchfactor\fontdimen3\font minus
  \fontdimen4\font\relax}
\providecommand{\BIBforeignlanguage}[2]{{%
\expandafter\ifx\csname l@#1\endcsname\relax
\typeout{** WARNING: IEEEtran.bst: No hyphenation pattern has been}%
\typeout{** loaded for the language `#1'. Using the pattern for}%
\typeout{** the default language instead.}%
\else
\language=\csname l@#1\endcsname
\fi
#2}}
\providecommand{\BIBdecl}{\relax}
\BIBdecl

\bibitem{Bergmann2019}
P.~Bergmann, M.~Fauser, D.~Sattlegger, and C.~Steger, ``{MVTec AD — A
  Comprehensive Real-World Dataset for Unsupervised Anomaly Detection},''
  \emph{Cvpr 2019}, pp. 9592--9600, 2019.

\bibitem{Pimentel2014}
\BIBentryALTinterwordspacing
M.~A. Pimentel, D.~A. Clifton, L.~Clifton, and L.~Tarassenko, ``{A review of
  novelty detection},'' \emph{Signal Processing}, vol.~99, pp. 215--249, 2014.
  [Online]. Available: \url{http://dx.doi.org/10.1016/j.sigpro.2013.12.026}
\BIBentrySTDinterwordspacing

\bibitem{Seebock2019}
P.~Seebock, J.~I. Orlando, T.~Schlegl, S.~M. Waldstein, H.~Bogunovic,
  S.~Klimscha, G.~Langs, and U.~Schmidt-Erfurth, ``{Exploiting Epistemic
  Uncertainty of Anatomy Segmentation for Anomaly Detection in Retinal OCT},''
  \emph{IEEE Transactions on Medical Imaging}, 2019.

\bibitem{Kendall2017}
A.~Kendall and Y.~Gal, ``{What uncertainties do we need in Bayesian deep
  learning for computer vision?}'' \emph{Advances in Neural Information
  Processing Systems}, vol. 2017-Decem, no. Nips, pp. 5575--5585, 2017.

\bibitem{Zhao2015}
H.~Zhao, O.~Gallo, I.~Frosio, and J.~Kautz, ``Loss functions for neural
  networks for image processing,'' \emph{arXiv preprint arXiv:1511.08861},
  2015.

\bibitem{Bergmann2019a}
P.~Bergmann, S.~L{\"{o}}we, M.~Fauser, D.~Sattlegger, and C.~Steger,
  ``{Improving unsupervised defect segmentation by applying structural
  similarity to autoencoders},'' \emph{VISIGRAPP 2019 - Proceedings of the 14th
  International Joint Conference on Computer Vision, Imaging and Computer
  Graphics Theory and Applications}, vol.~5, pp. 372--380, 2019.

\bibitem{Sabokrou2018}
M.~Sabokrou, M.~Khalooei, M.~Fathy, and E.~Adeli, ``{Adversarially Learned
  One-Class Classifier for Novelty Detection},'' \emph{Proceedings of the IEEE
  Computer Society Conference on Computer Vision and Pattern Recognition}, pp.
  3379--3388, 2018.

\bibitem{Schlegl2017}
\BIBentryALTinterwordspacing
T.~Schlegl, P.~Seeb{\"{o}}ck, S.~M. Waldstein, U.~Schmidt-Erfurth, and
  G.~Langs, ``{Unsupervised Anomaly Detection with Generative Adversarial
  Networks to Guide Marker Discovery},'' \emph{Lecture Notes in Computer
  Science (including subseries Lecture Notes in Artificial Intelligence and
  Lecture Notes in Bioinformatics)}, vol. 10265 LNCS, pp. 146--147, mar 2017.
  [Online]. Available: \url{http://arxiv.org/abs/1703.05921}
\BIBentrySTDinterwordspacing

\bibitem{Schlegl2019}
T.~Schlegl, P.~Seeb{\"{o}}ck, S.~M. Waldstein, G.~Langs, and
  U.~Schmidt-Erfurth, ``{f-AnoGAN: Fast unsupervised anomaly detection with
  generative adversarial networks},'' \emph{Medical Image Analysis}, vol.~54,
  no. January, pp. 30--44, 2019.

\bibitem{Akcay2019}
\BIBentryALTinterwordspacing
S.~Ak{\c{c}}ay, A.~Atapour-Abarghouei, and T.~P. Breckon, ``{Skip-GANomaly:
  Skip Connected and Adversarially Trained Encoder-Decoder Anomaly
  Detection},'' \emph{2019 International Joint Conference on Neural Networks
  (IJCNN)}, pp. 1--8, 2019. [Online]. Available:
  \url{http://arxiv.org/abs/1901.08954}
\BIBentrySTDinterwordspacing

\bibitem{Baur2019}
C.~Baur, B.~Wiestler, S.~Albarqouni, and N.~Navab, ``{Deep autoencoding models
  for unsupervised anomaly segmentation in brain MR images},'' \emph{Lecture
  Notes in Computer Science (including subseries Lecture Notes in Artificial
  Intelligence and Lecture Notes in Bioinformatics)}, vol. 11383 LNCS, pp.
  161--169, 2019.

\bibitem{Radford2015}
A.~Radford, L.~Metz, and S.~Chintala, ``Unsupervised representation learning
  with deep convolutional generative adversarial networks,'' \emph{arXiv
  preprint arXiv:1511.06434}, 2015.

\bibitem{Goodfellow1976}
I.~Goodfellow, J.~Pouget-Abadie, M.~Mirza, B.~Xu, D.~Warde-Farley, S.~Ozair,
  A.~Courville, and Y.~Bengio, ``{Generative Adversarial Nets},''
  \emph{Advances in neural information processing systems}, pp. 2672--2680,
  2014.

\bibitem{Snell2017}
J.~Snell, K.~Ridgeway, R.~Liao, B.~D. Roads, M.~C. Mozer, and R.~S. Zemel,
  ``Learning to generate images with perceptual similarity metrics,'' in
  \emph{2017 IEEE International Conference on Image Processing (ICIP)}.\hskip
  1em plus 0.5em minus 0.4em\relax IEEE, 2017, pp. 4277--4281.

\bibitem{Huang2019}
\BIBentryALTinterwordspacing
C.~Huang, J.~Cao, F.~Ye, M.~Li, Y.~Zhang, and C.~Lu, ``{Inverse-Transform
  AutoEncoder for Anomaly Detection},'' \emph{arXiv preprint arXiv:1911.10676},
  2019. [Online]. Available: \url{http://arxiv.org/abs/1911.10676}
\BIBentrySTDinterwordspacing

\bibitem{Chandola2009}
\BIBentryALTinterwordspacing
V.~Chandola, A.~Banerjee, and V.~Kumar, ``{Survey of Anomaly Detection},''
  \emph{ACM Computing Survey}, no. September, pp. 1--72, 2009. [Online].
  Available:
  \url{http://cucis.ece.northwestern.edu/projects/DMS/publications/AnomalyDetection.pdf}
\BIBentrySTDinterwordspacing

\bibitem{Napoletano2018}
P.~Napoletano, F.~Piccoli, and R.~Schettini, ``{Anomaly detection in
  nanofibrous materials by CNN-based self-similarity},'' \emph{Sensors
  (Switzerland)}, vol.~18, no.~1, 2018.

\bibitem{Staar2019}
\BIBentryALTinterwordspacing
B.~Staar, M.~L{\"{u}}tjen, and M.~Freitag, ``{Anomaly detection with
  convolutional neural networks for industrial surface inspection},''
  \emph{Procedia CIRP}, vol.~79, no. January 2019, pp. 484--489, 2019.
  [Online]. Available:
  \url{https://linkinghub.elsevier.com/retrieve/pii/S2212827119302409}
\BIBentrySTDinterwordspacing

\bibitem{Zhou2017}
C.~Zhou and R.~C. Paffenroth, ``{Anomaly Detection with Robust Deep
  Autoencoders},'' \emph{Proceedings of the 23rd ACM SIGKDD International
  Conference on Knowledge Discovery and Data Mining}, pp. 665--674, 2017.

\bibitem{Dehaene2020}
\BIBentryALTinterwordspacing
D.~Dehaene, O.~Frigo, S.~Combrexelle, and P.~Eline, ``{Iterative energy-based
  projection on a normal data manifold for anomaly localization},'' pp. 1--17,
  2020. [Online]. Available: \url{http://arxiv.org/abs/2002.03734}
\BIBentrySTDinterwordspacing

\bibitem{Haselmann2019}
M.~Haselmann, D.~P. Gruber, and P.~Tabatabai, ``{Anomaly Detection Using Deep
  Learning Based Image Completion},'' \emph{Proceedings - 17th IEEE
  International Conference on Machine Learning and Applications, ICMLA 2018},
  pp. 1237--1242, 2019.

\bibitem{Munawar2015}
A.~Munawar and C.~Creusot, ``{Structural inpainting of road patches for anomaly
  detection},'' \emph{Proceedings of the 14th IAPR International Conference on
  Machine Vision Applications, MVA 2015}, pp. 41--44, 2015.

\bibitem{Mei2018}
S.~Mei, Y.~Wang, and G.~Wen, ``{Automatic fabric defect detection with a
  multi-scale convolutional denoising autoencoder network model},''
  \emph{Sensors (Switzerland)}, vol.~18, no.~4, pp. 1--18, 2018.

\bibitem{Ronneberger2015}
O.~Ronneberger, P.~Fischer, and T.~Brox, ``{U-net: Convolutional networks for
  biomedical image segmentation},'' \emph{Lecture Notes in Computer Science
  (including subseries Lecture Notes in Artificial Intelligence and Lecture
  Notes in Bioinformatics)}, vol. 9351, pp. 234--241, 2015.

\bibitem{Zhong2020}
Z.~Zhong, L.~Zheng, G.~Kang, S.~Li, and Y.~Yang, ``{Random Erasing Data
  Augmentation},'' \emph{Proceedings of the AAAI Conference on Artificial
  Intelligence}, vol.~34, no.~07, pp. 13\,001--13\,008, 2020.

\bibitem{Fong2019}
R.~Fong and A.~Vedaldi, ``{Occlusions for effective data augmentation in image
  classification},'' \emph{Proceedings - 2019 International Conference on
  Computer Vision Workshop, ICCVW 2019}, pp. 4158--4166, 2019.

\bibitem{Liu2018}
G.~Liu, F.~A. Reda, K.~J. Shih, T.~C. Wang, A.~Tao, and B.~Catanzaro, ``{Image
  Inpainting for Irregular Holes Using Partial Convolutions},'' \emph{Lecture
  Notes in Computer Science (including subseries Lecture Notes in Artificial
  Intelligence and Lecture Notes in Bioinformatics)}, vol. 11215 LNCS, pp.
  89--105, 2018.

\end{thebibliography}
